\documentclass[aps,prd,superscriptaddress,showpacs,nofootinbib,notitlepage,twocolumn]{revtex4-1}
\usepackage{graphicx,subfigure,amsmath,amsfonts,amssymb,multirow,mathrsfs}
\usepackage{ulem}
\usepackage[colorlinks,linkcolor=blue,citecolor=blue,urlcolor=blue]{hyperref}
\usepackage[dvipsnames,svgnames]{xcolor}
\def\apj{ApJ~}                 
\def\apjl{ApJL~}                
\def\aap{A\&A~}                

\def\mnras{MNRAS~}             
\def\prd{Phys~Rev~D~}        
\def\prl{Phys~Rev~Lett~}    
\def\nat{Nature~}              

\def\physrep{Phys~Rep~}   

\renewcommand\apjl{ApJL}

\graphicspath{{./}{figures/}}

\begin{document}


\title{Aligned Hierarchical Black Hole Mergers in Active-Galactic-Nuclei Disks Revealed by GWTC-4}

\author{Yin-Jie Li$^\dagger$}
\affiliation{Key Laboratory of Dark Matter and Space Astronomy, Purple Mountain Observatory, Chinese Academy of Sciences, Nanjing 210023, People's Republic of China}

\author{Yuan-Zhu Wang$^\dagger$}
\affiliation{Institute for Theoretical Physics and Cosmology, Zhejiang University of Technology, Hangzhou, 310032, People's Republic of China}

\author{Shao-Peng Tang}
\affiliation{Key Laboratory of Dark Matter and Space Astronomy, Purple Mountain Observatory, Chinese Academy of Sciences, Nanjing 210023, People's Republic of China}

\author{Yi-Zhong Fan}
\email{The corresponding author: yzfan@pmo.ac.cn\\
$^\dagger$Contributed equally}
\affiliation{Key Laboratory of Dark Matter and Space Astronomy, Purple Mountain Observatory, Chinese Academy of Sciences, Nanjing 210023, People's Republic of China}
\affiliation{School of Astronomy and Space Science, University of Science and Technology of China, Hefei, Anhui 230026, People's Republic of China}

\begin{abstract}
The active galactic nucleus (AGN) accretion disks are ideal sites for hierarchical black hole (BH) mergers.
To robustly probe such a possibility,
we analyze binary black hole mergers in the GWTC-4 with a flexible mixture population model for component masses, spin magnitudes, and spin tilt angles, and identify two distinct subpopulations. 
In the second subpopulation characterized by high spin magnitudes $\chi\sim 0.8$ as well as the broad mass distribution up to $\gtrsim  150M_\odot$, 
we find a pronounced preference for spins aligned with the orbital angular momentum: an isotropic tilt distribution is strongly disfavored (logarithmic Bayes factor = 4.5).
The aligned events account for $\sim 0.57^{+0.23}_{-0.31}$ of the second subpopulation, corresponding to a local rate of $\sim 0.25^{+0.38}_{-0.16} ~ {\rm Gpc}^{-3} {\rm yr}^{-1}$  (all values reflect central 90\% credible intervals).  
These notable features naturally arise from hierarchical mergers embedded in AGN disks, where gas torques may effectively align spins. Our results suggest that AGN-disk hierarchical assembly may be one important channel for the present gravitational-wave sample, 
and provide concrete, testable predictions for future detection.
\end{abstract}
\maketitle

\textbf{\textit{Introduction.}} \label{sec:intro}
Thanks to the successful performance of the ground-based gravitational-wave detectors, in particular the advanced LIGO \citep{2015CQGra..32g4001L} and Virgo \citep{2015CQGra..32b4001A}, more than 180 gravitational wave events, most involving stellar-mass black hole pairs, have been officially reported \citep{2024PhRvD.109b2001A,2023PhRvX..13d1039A,2025arXiv250818082T}. Due to the absence of electromagnetic counterparts and the relatively poor localizations, one of the key open questions for the binary black hole merger (BBH) events is where and how they were formed  \citep{2022PhR...955....1M}.
The BBH formation channels can be broadly categorized into two main pathways: isolated binary evolution and dynamical formation in dense stellar environments like star clusters and around active galactic nuclei (AGN) disks.
Since different evolutionary channels yield BBHs with distinct characteristics, such as merger rate histories, spin properties, and mass distributions, the population analysis can play a crucial role in revealing  the formation pathways of BBHs \citep{2022PhR...955....1M,2021NatAs...5..749G}.
For instance, the BHs resulting from core collapses generally have low spins \citep{2019ApJ...881L...1F}, although they can be moderately spun up through binary interactions \citep{2018A&A...616A..28Q}. In contrast, second or higher generation BHs, which are remnants of previous mergers, exhibit notably different spin-magnitude distributions, typically peaking around $\sim 0.7$ \citep{2017ApJ...840L..24F, 2021NatAs...5..749G} or even higher for the aligned mergers \citep{2024A&A...685A..51V}. 
Indication for hierarchical mergers was suggested with GWTC-2 data under the assumption of star cluster channels \citep{2021ApJ...915L..35K}, and robust evidence was reported by resolving two distinct mass-spin subpopulaions among GWTC-3 events
without such a limitation \citep{2022ApJ...941L..39W, 2024PhRvL.133e1401L, 2024A&A...692A..80P}.
One subpopulation, characterized by low BH masses and spins ($\chi \lesssim 0.4$), aligns with stellar core-collapse origins. The other subpopulation, with higher masses and spins ($\chi \sim 0.75$), is indicative of hierarchical mergers. The hosts of these hierarchical mergers, however, remain an unresolved issue \citep{2024PhRvL.133e1401L,2025PhRvL.134a1401A,2025ApJ...987...65L}. For a few exceptional events, such as GW190521 and GW231123, the AGN disk-driven hierarchical merger origin is likely favored \citep{2020PhRvL.124y1102G,2026ApJ...999..127L}, but alternative reasonable explanations are still available \citep{2020ApJ...900L..13A,2023NatAs...7...11G,2025ApJ...993L..25A,2025ApJ...994L..37K}. As such, confirming the existence of a population of AGN disk-driven hierarchical mergers is an ongoing endeavor, which is the main goal of this work.

The distributions of spin orientations 
carry fundamental information on the formation mechanisms and environments of BBHs that lead to mergers. BBHs originating from isolated field binaries usually exhibit nearly aligned spins, whereas those formed through dynamic processes in star clusters have isotropic spins \citep{2022PhR...955....1M}. BBHs dynamically assembled within AGN disks, however, may show oriented spins (aligned or anti-aligned) due to the interaction with the AGN disk \citep{2019PhRvL.123r1101Y}.  
It has been proposed that counter-alignment of BHs' spins serves as a key discriminant between the AGN disk scenario and other pathways, such as isolated binary evolution or dynamical assembly in clusters \citep{2021NatAs...5..749G}. However, this distinction may be challenged by some critical factors, for example,  anti-aligned (or retrograde) BBHs could be preferentially ionized as they spend
 more time at wider separation \citep{2021ApJ...923L..23W}, and the dynamical encounters may flip their spins to align (or prograde) configurations \citep{2022Natur.603..237S,2024MNRAS.531.3479M}. 
Nevertheless, the joint analysis of component masses, spin magnitudes, and spin orientations could provide a compelling
signature of AGN disk-driven hierarchical mergers \citep{2021NatAs...5..749G,2024MNRAS.531.3479M}.

\begin{figure*}
	\centering  
\includegraphics[width=0.98\linewidth]{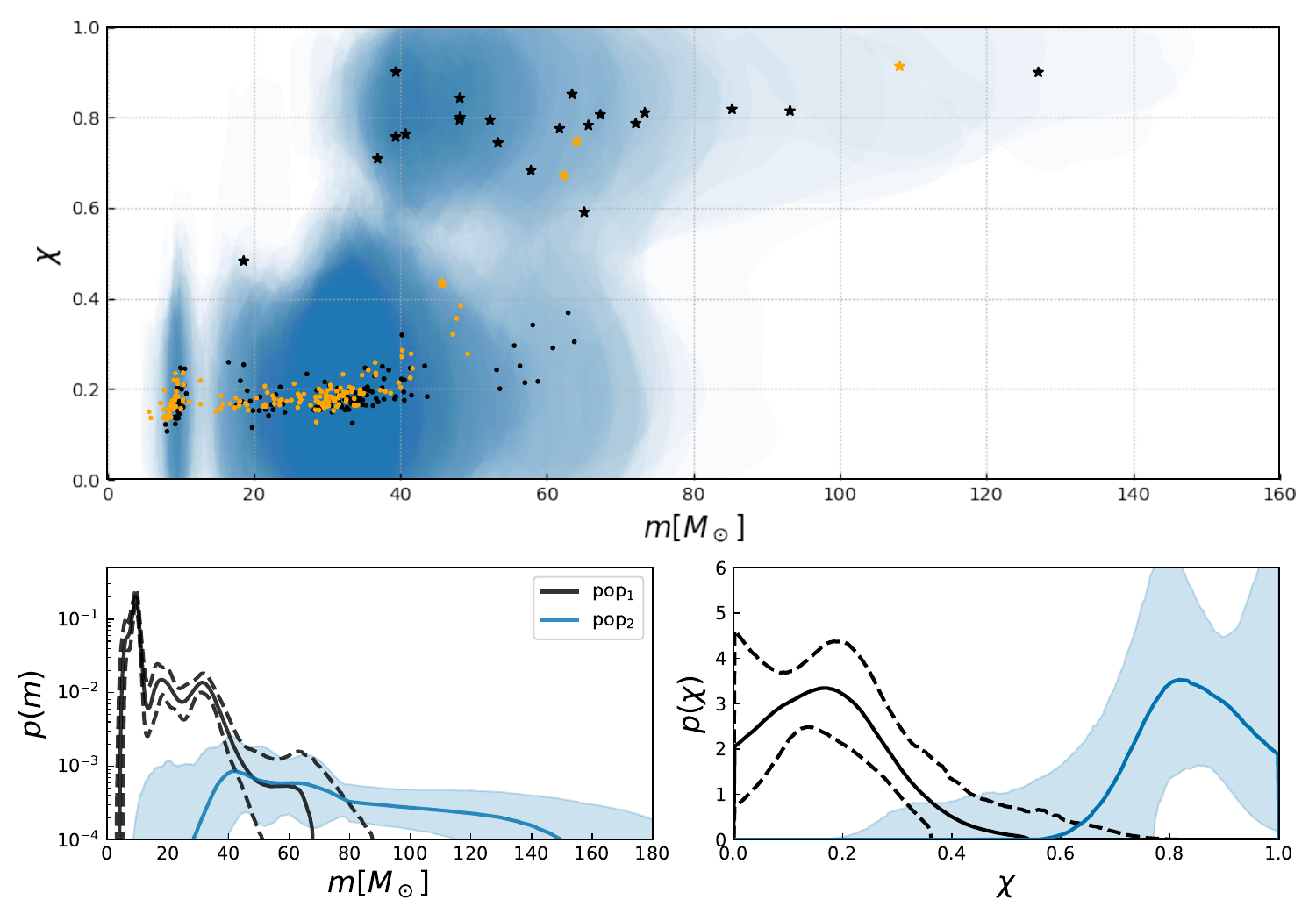}
\caption{Top: Posteriors of individual component masses and spin magnitudes of BBHs in GWTC-4 reweighed to a population-informed prior inferred by our nonparametric model. The shaded areas mark the 90\% credible regions and the black (orange) points stand for the mean values for the primary (secondary) BHs. The BHs with mean values of $\chi$ $>0.4$ are marked with stars. 
Bottom: Reconstructed component mass (left) and spin magnitude (right) distributions of the first and second subpopulations; the solid curves are the medians and the dashed curves are for the 90\% credible intervals.}
\label{fig:spin}
\end{figure*}

\textbf{\textit{Approaches.}} \label{sec:method}
Previous studies have investigated the origins of hierarchical mergers with effective spins in GW data, yet the evidence remains  inconclusive  \citep{2021arXiv211010838W,2025PhRvL.134a1401A,2025ApJ...987...65L,2025ApJ...990..217M}.
More frustratingly, AGN disk-driven hierarchical mergers may be difficult to distinguish significantly from some field evolution channels, from the perspective of effective spins \citep{2021ApJ...910..152Z}.
Therefore, in this work, we conduct a population analysis of the newly released GWTC-4 data \cite{2025arXiv250818082T}, involving spin magnitudes, spin orientations, and component masses of the BBH events,
using a novel nonparametric model. 

Following \citep{2025arXiv250818083T}, we select 153 BBH events for analysis, using a false-alarm rate threshold $< 1~{\rm yr}^{-1}$ in GWTC-4, and employ the common hierarchical Bayesian inference \citep{2019MNRAS.486.1086M,2025arXiv250818083T} to estimate the population properties of BBHs. The detailed formulation is provided in Section~\uppercase\expandafter{\romannumeral1} of Supplemental Material \citep{supp} 

The population model builds upon our semi-parametric framework developed in \citet{2024PhRvL.133e1401L}, in which only the component-mass distribution was modeled flexibly since the data sample was relatively small. Benefited from the significantly increased number of the BBH events, we are able to take a much more flexible mixture population model for component masses, spin magnitudes, and spin–orbit tilts, which reads 
\begin{equation}
\begin{aligned}
\pi(\lambda|\Lambda)& \propto \pi(m_1,\chi_1,\cos\theta_1|\Lambda)\times \pi(m_2,\chi_2,\cos\theta_2|\Lambda)\\
& \times P_{\rm pair}(q|\Lambda) \times \pi_{\rm z}(z|\gamma),
\end{aligned}
\end{equation}
note that
\begin{equation}\label{twopop}
\begin{aligned}
&\pi(m,\chi,\cos\theta|\Lambda)= \\
&\sum_{i=1,2} P_{\mathcal{PS}}(m|\Lambda_i) \times P_{\mathcal{S}}(\chi |\Lambda_i) \times P_{\mathcal{S}}(\cos\theta |\Lambda_i) \times r_i,
\end{aligned}
\end{equation}
where $P_{\mathcal{PS}}$ and $P_{\mathcal{S}}$ are the $PowerLaw+Spline$ and $Spline$ models for the distributions of component masses and spin properties. 
Besides $P_{\mathcal{PS}}(m|\Lambda_i)$, now the distributions of spin magnitudes $P_{\mathcal{S}}(\chi |\Lambda_i)$, spin orientations $P_{\mathcal{S}}(\cos\theta |\Lambda_i)$, and the pairing function $P_{\rm pair}(q|\Lambda)$ are also described by flexible, data-driven representations, i.e., exponentiated cubic splines $p(x)\propto e^{f(x)}$,
please see Section~\uppercase\expandafter{\romannumeral2} of Supplemental Material \citep{supp} for the details of these sub-models. 
Unlike the \textsc{B-Spline}-based model adopted by \citet{2025arXiv250818083T} and \citet{2023ApJ...946...16E}, which focuses on marginal distributions of individual parameters, our mixture model captures intrinsic correlations among parameters and enables the identification of potential astrophysical subpopulations.

\textbf{\textit{Results.}} \label{sec:result}
Our analysis unambiguously identified two subpopulations of BHs with a Bayes factor of $\ln\mathcal{B}=12$, considerably strengthening our earlier findings based on GWTC-3 \citep{2022ApJ...941L..39W,2024PhRvL.133e1401L}. Note that some recent studies based on GWTC-4 \citep{2026Natur.652..874T,2025arXiv250904637A,2025arXiv250909123A,2025ApJ...994..261A,2025arXiv250915646B} have also identified two subpopulations, though these approaches are not as flexible as ours and the resulting mass cutoff of the first subpopulation is lower than what we find here. Below we focus on our findings. Figure~\ref{fig:spin} shows the component-mass and spin-magnitude distributions of two subpopulations.
The first subpopulation exhibits spin magnitudes $\lesssim 0.4$ peaking at $\sim 0.2$, and masses $\lesssim m_{\rm max,1}=66.7_{-8.2}^{+11.4} M_{\odot}$ (68\% highest density interval), consistent with first-generation BHs. The second subpopulation, with spin magnitudes $\gtrsim 0.5$ peaking at $\sim 0.8$ and populating the expected pair instability mass gap, matches the properties of the remnants of previous mergers \citep{2017PhRvD..95l4046G,2017ApJ...840L..24F,2021NatAs...5..749G}. The two subpopulations are clearly separated, as illustrated by the distributions of spin magnitudes and component masses in Figure~\ref{fig:spin}. We note that the minimum mass of the high-spin subpopulation could be $\lesssim 20 M_{\odot}$. One interesting candidate is GW231118\_005626, of which primary object is consistent with being the second generation BH resulting from a previous merger, with component masses in the $\sim 9 - 10 M_\odot$ range, suggesting that {\it some BBHs with masses peaking at $\sim 10M_\odot$ may be dynamically assembled}.  Furthermore, the maximum mass of the low-spin subpopulation may reach $\gtrsim 65M_\odot$ {for a reasonable  $S$ factor
of the $^{12}C(\alpha,\gamma)^{16}O$ radiative capture (see \citep{Wang:2025nhf} for a dedicated investigation)}, 
while is still in line with the stellar evolution theories \citep{2021ApJ...912L..31W}.
The fraction of hierarchical mergers (containing at least one higher-generation BH) is $2.5_{-1.3}^{+3.2}\%$ corresponding to a  merger rate of $0.47_{-0.25}^{+0.58}~{\rm Gpc}^{-3}~{\rm yr}^{-1}$, consistent with that inferred from GWTC-3 \citep{2024PhRvL.133e1401L}. 

\begin{figure}
	\centering  
\includegraphics[width=0.99\linewidth]{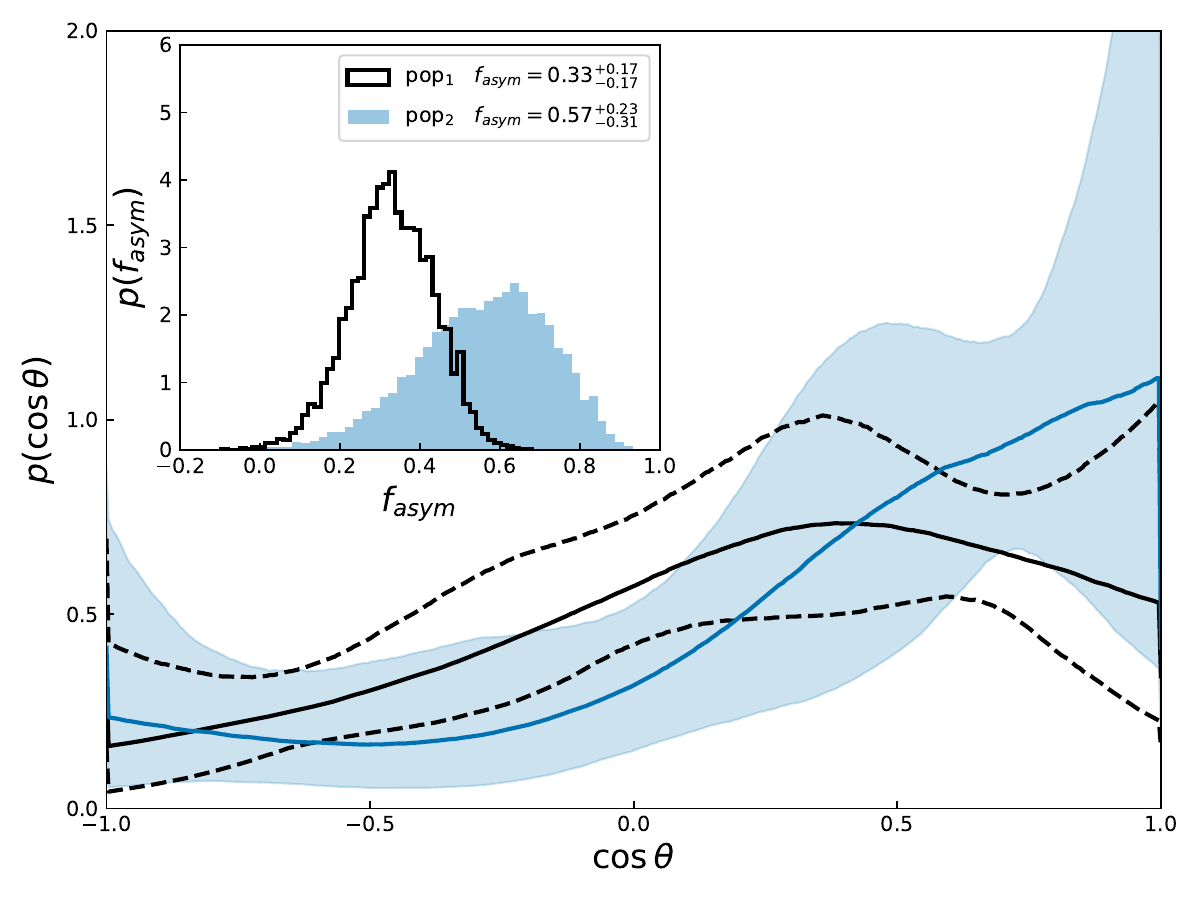}
\caption{Main: Reconstructed spin orientation distributions of the first and second subpopulations. Insert: The asymmetric fraction of $\cos\theta$ distribution for the first and second subpopulation the values are for median and 90\% credible intervals.}
\label{fig:asym}
\end{figure}

Our main discovery is that neither subpopulation has an isotropic spin distribution (see Figure \ref{fig:asym}). The asymmetric fraction of the $\cos\theta$ distribution is defined as 
\begin{equation*}
f_{\rm asym}=\int_{0}^{1}{[P(\cos\theta)-P(-\cos\theta)]}\mathrm{d}\cos\theta.
\end{equation*}
The symmetric fraction is hence $f_{\rm sym}=1-|f_{\rm asym}|$. In principle, the isotropic (i.e.,  $\cos\theta$ distributes uniformly in [-1,1]) fraction $f_{\rm iso}$ should be $\lesssim f_{\rm sym}$, so that the non-isotropic fraction $f_{\rm non-iso}=1-f_{\rm iso}$ should be $\gtrsim |f_{\rm asym}|$. 

For the first subpopulation, we have $f_{\rm asym}=0.33^{+0.17}_{-0.18}$, and an isotropic distribution is disfavored at a confidence level of $>99.87\%$.  
For the BHs with a stellar collapse origin, $f_{\rm non-iso} \gtrsim 0.33^{+0.17}_{-0.18}$ can be attributed to the field binary evolution channels
\citep{2016ApJ...832L...2R,2021ApJ...910..152Z,2022PhR...955....1M}, and the underlying fraction could be larger, considering a mass dependent $\cos\theta$ distribution \citep{2022ApJ...941L..39W,2024ApJ...977...67L}.

For the second subpopulation, it turns out that an isotropic distribution  is strongly disfavored compared to a distribution with a preference for alignment, by a Bayes factor of $\ln\mathcal{B}=4.5$, see Section~\uppercase\expandafter{\romannumeral3} of Supplemental Material \citep{supp} for the detailed model comparison. For formation channels in star clusters, an $f_{\rm asym}\sim0$ is expected \citep{2025PhRvL.134a1401A}, whereas in AGN disks, an $f_{\rm asym}>0$ ($f_{\rm asym}<0$ for anti-alignment) is plausible \citep{2017PhRvD..95l4046G,2017ApJ...840L..24F,2019PhRvL.123r1101Y}. Our inference yields $f_{\rm asym} = 0.57^{+0.23}_{-0.31}$, with $f_{\rm asym} > 0$ at the 99.85\% credible level, as depicted in Figure~\ref{fig:asym}. This implies that formation channels leading to hierarchical mergers with isotropic spins, e.g., star-cluster channels \citep{2016ApJ...832L...2R, 2022PhR...955....1M}, account for a fraction $f_{\rm iso}\lesssim 0.43^{+0.31}_{-0.23}$.
We infer such non-isotropic fraction to be $f_{\rm non-iso}\gtrsim 0.57^{+0.23}_{-0.31}$, which is consistent with the non-isotropic fraction $\zeta_2=0.75^{+0.22}_{-0.36}$ obtained independently with a parametric model in Section~\uppercase\expandafter{\romannumeral3} of Supplemental Material \citep{supp}. It is particularly noteworthy that the mergers of aligned BBHs in the AGN disks can also help account for the very high $\chi\gtrsim 0.8$ (see Figure~\ref{fig:spin}) that is difficult to explain well in other scenarios 
\citep{2021NatAs...5..749G,2024A&A...685A..51V}.

\begin{figure*}
	\centering  
\includegraphics[width=0.61\linewidth]{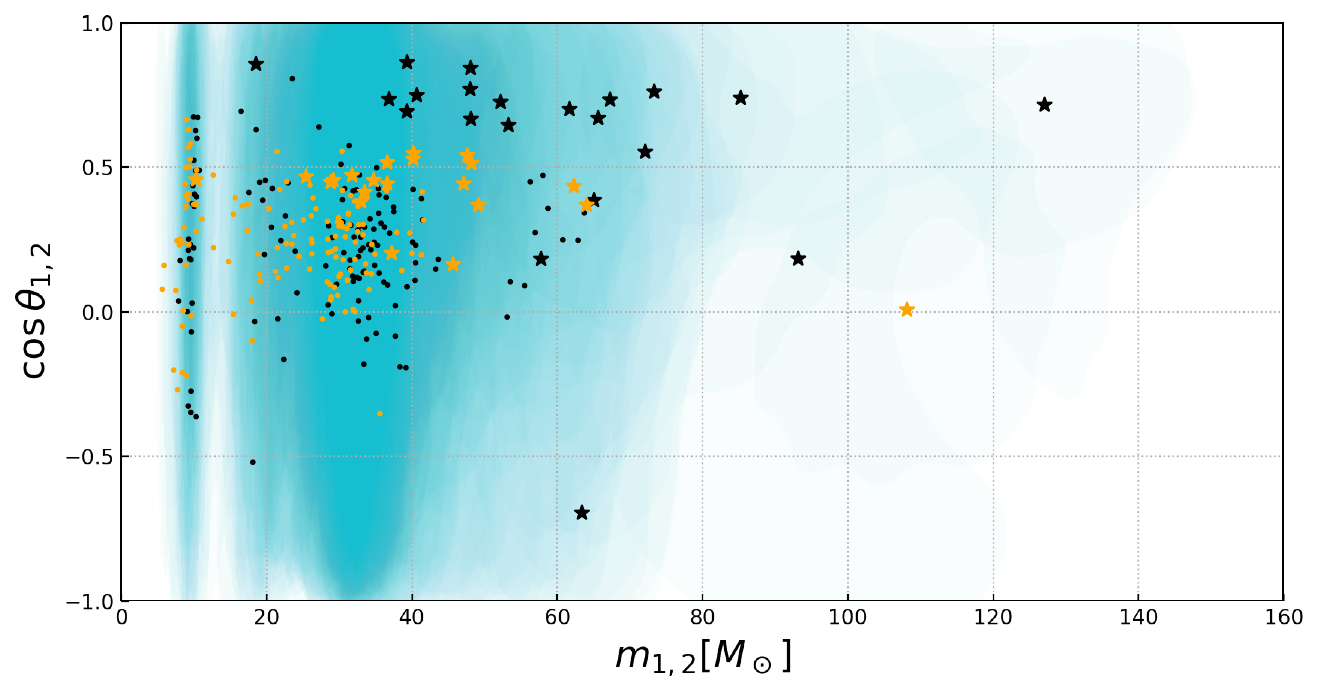}
\includegraphics[width=0.34\linewidth]{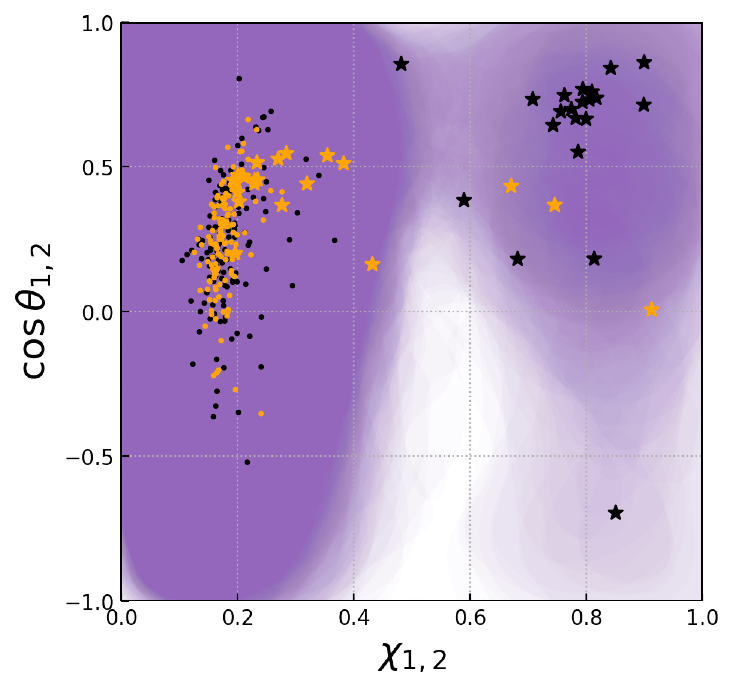}
\includegraphics[width=0.61\linewidth]{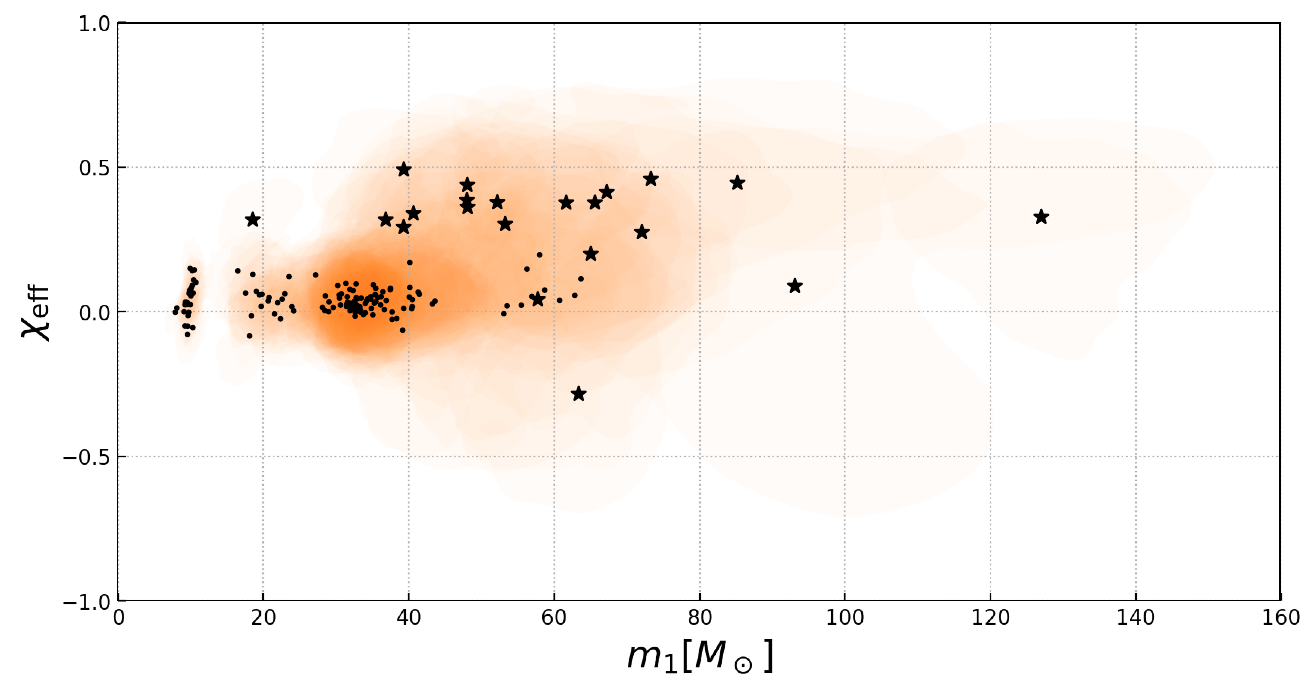}
\includegraphics[width=0.34\linewidth]{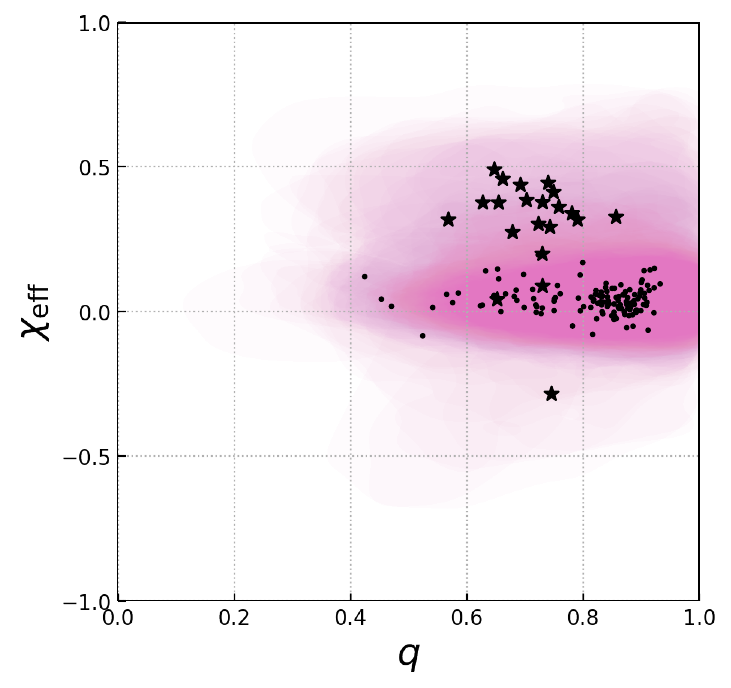}
\caption{Population-informed posterior distribution for each event. The shaded areas mark the 90\% credible regions and the black (orange) points stand for the mean values for the primary (secondary) BHs. The BHs (BBHs) with mean values of $\chi$ ($\chi_1$) $>0.4$ are marked with stars. Note that, in the top panels, the secondary BHs are also marked if the corresponding primary with $\bar{\chi_1}$ $>0.4$. It is clear that $\cos\theta$ ($\chi_{\rm eff}$) of the BHs (BBHs) in the second subpopulation with high masses and high spins $\chi \sim 0.5 - 0.9$ exhibit a significant preference for positive values.}
\label{fig:sample}
\end{figure*}

Figure~\ref{fig:sample} displays the population-informed posterior distribution for each event, with (potential) hierarchical mergers (containing high-spin BHs) highlighted with stars. In the top panels, we observe that the high-spin and high-mass subpopulation favors aligned systems (positive $\cos\theta$), a trend that becomes more pronounced in the $\chi-\cos\theta$ panel, where the two subpopulations are distinctly separated. In the bottom panels, we display the distributions of effective spins $\chi_{\rm eff}\equiv(\cos\theta_1\chi_1+\cos\theta_2\chi_2q)/(1+q)$. The broad and asymmetric $\chi_{\rm eff}$ distribution (skewed towards positive values) of the second subpopulation is evident, while the first subpopulation shows a narrower $\chi_{\rm eff}$ distribution. These results are in concordance with independent analysis on the $m_1-q-\chi_{\rm eff}$ distribution \citep{2021arXiv211010838W,2025ApJ...987...65L}, and the marginalized $\chi_{\rm eff}$ distribution \citep{2025arXiv250818083T}. Remarkably, the $\chi_{\rm eff}$ distribution of the high-spin subpopulation aligns well with that predicted in  the hierarchical mergers in AGN disks 
\citep{2019PhRvL.123r1101Y,2025ApJ...990..217M}, in agreement with our argument basing on $f_{\rm asym}$ presented above.
{Some works suggest that the mass-ratio distribution for the AGN-driven channels may peak at a low value (e.g. $\sim 0.2$) \citep{2019ApJ...876..122Y,2021ApJ...920L..42G}. However, we do not find such events in our samples. There are two possible reasons. One is that the binaries generated in \citet{2019ApJ...876..122Y} are assumed to be randomly paired, while the real pairing mechanism may have preference for symmetric systems, as already noticed by these authors. The other is that given a (higher-generation) primary BH, a lower mass ratio required a lighter secondary BH, making the binary harder to be detected.
}

\textbf{\textit{Discussion.}} \label{sec:summary}
Our analysis identifies a significant excess of aligned systems in the second subpopulation, which we would attribute to the mergers driven by AGN disks. 
In some literature, it was also suggested that there may be an excess of anti-aligned systems in the AGN disk \citep{2021NatAs...5..749G}, but our search finds no evidence for such a group in the current BBH sample. 
We find the fraction of anti-aligned assembly is consistent with zero, see Section~\uppercase\expandafter{\romannumeral3}~B of Supplemental Material \citep{supp}.
The absence of anti-aligned assembly 
likely points towards
the effective gas hardening in the AGN disks \citep{2024A&A...685A..51V}, 
since the gas-drag helps align both binary orbital angular momentum and the BH spins to the AGN disks.

For the suggested AGN disk-driven hierarchical merger scenario, we estimate the rate to be $R_{\rm AGN,HM}(z=0) \sim 0.25_{-0.16}^{+0.38} ~{\rm Gpc}^{-3} {\rm yr}^{-1}$. It was argued in \citet{2019PhRvL.123r1101Y} that roughly half of AGN disk mergers are hierarchical, we hence estimate a total AGN channel merger rate of $R_{\rm AGN}(z=0) \sim 0.50_{-0.32}^{+0.75} ~ {\rm Gpc}^{-3} {\rm yr}^{-1}$, which is $2.7_{-1.8}^{+4.0}\%$ of the local BBH merger rate \citep{2025arXiv250818083T}, 
corresponding to $f_{\rm AGN,Obs.}\sim 10.9_{-6.4}^{+7.4}\%$ of the observed sample, thanks to the stronger gravitational wave radiation of the AGN disk-driven BBH events. An upper limit of $f_{\rm AGN,Obs.} < 19.6_{-6.3}^{+9.1}\%$ is set with the total hierarchical merger rate if some BBH events are misaligned due to for instance the absence of gas hardening in some low-accretion-rate AGNs.
 Our $f_{\rm AGN,Obs.}$ is comparable with the value inferred via the spatial correlation analysis of the BBH events and the AGNs \citep{2025ApJ...989L..15Z}. More intriguingly, after the submission of this work, the detection of unexpected leading delays in broad H$\beta$ line reverberations in the quasar PHL 1092 was reported, and the presence of multiple stellar-mass BHs in AGN disks is preferred
\citep{2025arXiv251107716W}, which lends support to our AGN-driven disk channel interpretation of the aligned hierarchical black holes identified in this work.

Though the uncertainties of the spin measurements of BH components are still relatively large, a group of massive BHs with a high spin $\chi \geq 0.8$ likely emerges (see Figure~\ref{fig:spin}). Such a high $\chi$ is not expected in the hierarchical mergers with the random orientations \citep{2021NatAs...5..749G}. A very high $\chi$, however, is natural for the mergers of the aligned BBHs due to the effective gas hardening in the AGN disks \citep{2024A&A...685A..51V} (though see \citep{2025ApJ...994L..37K,2026ApJ...996L..44B} for alternative possibilities).
One additional prediction of such a  specific scenario
is the smooth extension of the maximum mass to $\sim 10^{4}M_\odot$, which will be directly tested by the observations of the BBH mergers with the future space-based gravitational wave detectors. 

In both the first and second subpopulations, spin orientations are non-isotropic, suggesting that star clusters do not dominate the formation channels for either first-generation or higher-generation mergers.
However, discerning whether AGN disks significantly contribute to first-generation BBHs remains challenging. Because traditional field evolution channels may also yield BBHs with aligned spins \citep{2016ApJ...832L...2R}, and a subpopulation from field evolution is likely emerging \citep{2022ApJ...941L..39W,2023arXiv230401288G,2024ApJ...977...67L}. Examining the spatial correlation between GW sources and their host environments may help resolve this issue \citep{2025ApJ...989L..15Z}.
The mass-dependent $\cos\theta$ distribution observed in the second subpopulation could arise because star clusters and AGN disks are predicted to produce different mass distributions \citep{2025PhRvD.112f3034X,2024A&A...685A..51V}. This characteristic might assist in distinguishing between subpopulations of AGN-driven and cluster-driven hierarchical mergers as the dataset grows in the future. 

\begin{acknowledgments}
This work is supported in part by the National Natural Science Foundation of China (No. 12233011, No. 12588101, No. 12503059, No. 12203101, No. 12303056),  the New Cornerstone Science Foundation through the XPLORER PRIZE, the General Fund (No. 2024M753495) of the China Postdoctoral Science Foundation, and the Priority Research Program of the Chinese Academy of Sciences (No. XDB0550400). This research has made use of data and software obtained from the Gravitational Wave Open Science Center (https://www.gw-openscience.org), a service of LIGO Laboratory, the LIGO Scientific Collaboration and the Virgo Collaboration. LIGO is funded by the U.S. National Science Foundation. Virgo is funded by the French Centre National de Recherche Scientifique (CNRS), the Italian Istituto Nazionale della Fisica Nucleare (INFN) and the Dutch Nikhef, with contributions by Polish and Hungarian institutes.
\end{acknowledgments}

\clearpage

\bibliographystyle{apsrev4-1}
\bibliography{ref.bib}

\begin{thebibliography}{64}%
\makeatletter
\providecommand \@ifxundefined [1]{%
 \@ifx{#1\undefined}
}%
\providecommand \@ifnum [1]{%
 \ifnum #1\expandafter \@firstoftwo
 \else \expandafter \@secondoftwo
 \fi
}%
\providecommand \@ifx [1]{%
 \ifx #1\expandafter \@firstoftwo
 \else \expandafter \@secondoftwo
 \fi
}%
\providecommand \natexlab [1]{#1}%
\providecommand \enquote  [1]{``#1''}%
\providecommand \bibnamefont  [1]{#1}%
\providecommand \bibfnamefont [1]{#1}%
\providecommand \citenamefont [1]{#1}%
\providecommand \href@noop [0]{\@secondoftwo}%
\providecommand \href [0]{\begingroup \@sanitize@url \@href}%
\providecommand \@href[1]{\@@startlink{#1}\@@href}%
\providecommand \@@href[1]{\endgroup#1\@@endlink}%
\providecommand \@sanitize@url [0]{\catcode `\\12\catcode `\$12\catcode
  `\&12\catcode `\#12\catcode `\^12\catcode `\_12\catcode `\%12\relax}%
\providecommand \@@startlink[1]{}%
\providecommand \@@endlink[0]{}%
\providecommand \url  [0]{\begingroup\@sanitize@url \@url }%
\providecommand \@url [1]{\endgroup\@href {#1}{\urlprefix }}%
\providecommand \urlprefix  [0]{URL }%
\providecommand \Eprint [0]{\href }%
\providecommand \doibase [0]{http://dx.doi.org/}%
\providecommand \selectlanguage [0]{\@gobble}%
\providecommand \bibinfo  [0]{\@secondoftwo}%
\providecommand \bibfield  [0]{\@secondoftwo}%
\providecommand \translation [1]{[#1]}%
\providecommand \BibitemOpen [0]{}%
\providecommand \bibitemStop [0]{}%
\providecommand \bibitemNoStop [0]{.\EOS\space}%
\providecommand \EOS [0]{\spacefactor3000\relax}%
\providecommand \BibitemShut  [1]{\csname bibitem#1\endcsname}%
\let\auto@bib@innerbib\@empty
\bibitem [{\citenamefont {{LIGO Scientific Collaboration}}\ \emph
  {et~al.}(2015)\citenamefont {{LIGO Scientific Collaboration}}, \citenamefont
  {{Aasi}}, \citenamefont {{Abbott}}, \citenamefont {{Abbott}} \emph
  {et~al.}}]{2015CQGra..32g4001L}%
  \BibitemOpen
  \bibfield  {author} {\bibinfo {author} {\bibnamefont {{LIGO Scientific
  Collaboration}}}, \bibinfo {author} {\bibfnamefont {J.}~\bibnamefont
  {{Aasi}}}, \bibinfo {author} {\bibfnamefont {B.~P.}\ \bibnamefont
  {{Abbott}}}, \bibinfo {author} {\bibfnamefont {R.}~\bibnamefont {{Abbott}}},
  \emph {et~al.},\ }\href {\doibase 10.1088/0264-9381/32/7/074001} {\bibfield
  {journal} {\bibinfo  {journal} {Classical and Quantum Gravity}\ }\textbf
  {\bibinfo {volume} {32}},\ \bibinfo {eid} {074001} (\bibinfo {year}
  {2015})},\ \Eprint {http://arxiv.org/abs/1411.4547} {arXiv:1411.4547 [gr-qc]}
  \BibitemShut {NoStop}%
\bibitem [{\citenamefont {{Acernese}}\ \emph {et~al.}(2015)\citenamefont
  {{Acernese}}, \citenamefont {{Agathos}}, \citenamefont {{Agatsuma}},
  \citenamefont {{Aisa}} \emph {et~al.}}]{2015CQGra..32b4001A}%
  \BibitemOpen
  \bibfield  {author} {\bibinfo {author} {\bibfnamefont {F.}~\bibnamefont
  {{Acernese}}}, \bibinfo {author} {\bibfnamefont {M.}~\bibnamefont
  {{Agathos}}}, \bibinfo {author} {\bibfnamefont {K.}~\bibnamefont
  {{Agatsuma}}}, \bibinfo {author} {\bibfnamefont {D.}~\bibnamefont {{Aisa}}},
  \emph {et~al.},\ }\href {\doibase 10.1088/0264-9381/32/2/024001} {\bibfield
  {journal} {\bibinfo  {journal} {Classical and Quantum Gravity}\ }\textbf
  {\bibinfo {volume} {32}},\ \bibinfo {eid} {024001} (\bibinfo {year}
  {2015})},\ \Eprint {http://arxiv.org/abs/1408.3978} {arXiv:1408.3978 [gr-qc]}
  \BibitemShut {NoStop}%
\bibitem [{\citenamefont {{Abbott}}\ \emph {et~al.}(2024)\citenamefont
  {{Abbott}}, \citenamefont {{Abbott}}, \citenamefont {{Acernese}},
  \citenamefont {{Ackley}}, \citenamefont {{Adams}} \emph
  {et~al.}}]{2024PhRvD.109b2001A}%
  \BibitemOpen
  \bibfield  {author} {\bibinfo {author} {\bibfnamefont {R.}~\bibnamefont
  {{Abbott}}}, \bibinfo {author} {\bibfnamefont {T.~D.}\ \bibnamefont
  {{Abbott}}}, \bibinfo {author} {\bibfnamefont {F.}~\bibnamefont
  {{Acernese}}}, \bibinfo {author} {\bibfnamefont {K.}~\bibnamefont
  {{Ackley}}}, \bibinfo {author} {\bibfnamefont {C.}~\bibnamefont {{Adams}}},
  \emph {et~al.},\ }\href {\doibase 10.1103/PhysRevD.109.022001} {\bibfield
  {journal} {\bibinfo  {journal} {\prd}\ }\textbf {\bibinfo {volume} {109}},\
  \bibinfo {eid} {022001} (\bibinfo {year} {2024})},\ \Eprint
  {http://arxiv.org/abs/2108.01045} {arXiv:2108.01045 [gr-qc]} \BibitemShut
  {NoStop}%
\bibitem [{\citenamefont {{Abbott}}\ \emph
  {et~al.}(2023{\natexlab{a}})\citenamefont {{Abbott}}, \citenamefont
  {{Abbott}}, \citenamefont {{Acernese}}, \citenamefont {{Ackley}},
  \citenamefont {{Adams}} \emph {et~al.}}]{2023PhRvX..13d1039A}%
  \BibitemOpen
  \bibfield  {author} {\bibinfo {author} {\bibfnamefont {R.}~\bibnamefont
  {{Abbott}}}, \bibinfo {author} {\bibfnamefont {T.~D.}\ \bibnamefont
  {{Abbott}}}, \bibinfo {author} {\bibfnamefont {F.}~\bibnamefont
  {{Acernese}}}, \bibinfo {author} {\bibfnamefont {K.}~\bibnamefont
  {{Ackley}}}, \bibinfo {author} {\bibfnamefont {C.}~\bibnamefont {{Adams}}},
  \emph {et~al.},\ }\href {\doibase 10.1103/PhysRevX.13.041039} {\bibfield
  {journal} {\bibinfo  {journal} {Physical Review X}\ }\textbf {\bibinfo
  {volume} {13}},\ \bibinfo {eid} {041039} (\bibinfo {year}
  {2023}{\natexlab{a}})},\ \Eprint {http://arxiv.org/abs/2111.03606}
  {arXiv:2111.03606 [gr-qc]} \BibitemShut {NoStop}%
\bibitem [{\citenamefont {{The LIGO Scientific Collaboration}}\ \emph
  {et~al.}(2025{\natexlab{a}})\citenamefont {{The LIGO Scientific
  Collaboration}}, \citenamefont {{the Virgo Collaboration}}, \citenamefont
  {{the KAGRA Collaboration}} \emph {et~al.}}]{2025arXiv250818082T}%
  \BibitemOpen
  \bibfield  {author} {\bibinfo {author} {\bibnamefont {{The LIGO Scientific
  Collaboration}}}, \bibinfo {author} {\bibnamefont {{the Virgo
  Collaboration}}}, \bibinfo {author} {\bibnamefont {{the KAGRA
  Collaboration}}},  \emph {et~al.},\ }\href {\doibase
  10.48550/arXiv.2508.18082} {\bibfield  {journal} {\bibinfo  {journal} {arXiv
  e-prints}\ ,\ \bibinfo {eid} {arXiv:2508.18082}} (\bibinfo {year}
  {2025}{\natexlab{a}})},\ \Eprint {http://arxiv.org/abs/2508.18082}
  {arXiv:2508.18082 [gr-qc]} \BibitemShut {NoStop}%
\bibitem [{\citenamefont {{Mandel}}\ and\ \citenamefont
  {{Farmer}}(2022)}]{2022PhR...955....1M}%
  \BibitemOpen
  \bibfield  {author} {\bibinfo {author} {\bibfnamefont {I.}~\bibnamefont
  {{Mandel}}}\ and\ \bibinfo {author} {\bibfnamefont {A.}~\bibnamefont
  {{Farmer}}},\ }\href {\doibase 10.1016/j.physrep.2022.01.003} {\bibfield
  {journal} {\bibinfo  {journal} {\physrep}\ }\textbf {\bibinfo {volume}
  {955}},\ \bibinfo {pages} {1} (\bibinfo {year} {2022})},\ \Eprint
  {http://arxiv.org/abs/1806.05820} {arXiv:1806.05820 [astro-ph.HE]}
  \BibitemShut {NoStop}%
\bibitem [{\citenamefont {{Gerosa}}\ and\ \citenamefont
  {{Fishbach}}(2021)}]{2021NatAs...5..749G}%
  \BibitemOpen
  \bibfield  {author} {\bibinfo {author} {\bibfnamefont {D.}~\bibnamefont
  {{Gerosa}}}\ and\ \bibinfo {author} {\bibfnamefont {M.}~\bibnamefont
  {{Fishbach}}},\ }\href {\doibase 10.1038/s41550-021-01398-w} {\bibfield
  {journal} {\bibinfo  {journal} {Nature Astronomy}\ ,\ \bibinfo {pages} {749}}
  (\bibinfo {year} {2021})},\ \Eprint {http://arxiv.org/abs/2105.03439}
  {arXiv:2105.03439 [astro-ph.HE]} \BibitemShut {NoStop}%
\bibitem [{\citenamefont {{Fuller}}\ and\ \citenamefont
  {{Ma}}(2019)}]{2019ApJ...881L...1F}%
  \BibitemOpen
  \bibfield  {author} {\bibinfo {author} {\bibfnamefont {J.}~\bibnamefont
  {{Fuller}}}\ and\ \bibinfo {author} {\bibfnamefont {L.}~\bibnamefont
  {{Ma}}},\ }\href {\doibase 10.3847/2041-8213/ab339b} {\bibfield  {journal}
  {\bibinfo  {journal} {\apjl}\ }\textbf {\bibinfo {volume} {881}},\ \bibinfo
  {eid} {L1} (\bibinfo {year} {2019})},\ \Eprint
  {http://arxiv.org/abs/1907.03714} {arXiv:1907.03714 [astro-ph.SR]}
  \BibitemShut {NoStop}%
\bibitem [{\citenamefont {{Qin}}\ \emph {et~al.}(2018)\citenamefont {{Qin}},
  \citenamefont {{Fragos}}, \citenamefont {{Meynet}}, \citenamefont
  {{Andrews}}, \citenamefont {{S{\o}rensen}},\ and\ \citenamefont
  {{Song}}}]{2018A&A...616A..28Q}%
  \BibitemOpen
  \bibfield  {author} {\bibinfo {author} {\bibfnamefont {Y.}~\bibnamefont
  {{Qin}}}, \bibinfo {author} {\bibfnamefont {T.}~\bibnamefont {{Fragos}}},
  \bibinfo {author} {\bibfnamefont {G.}~\bibnamefont {{Meynet}}}, \bibinfo
  {author} {\bibfnamefont {J.}~\bibnamefont {{Andrews}}}, \bibinfo {author}
  {\bibfnamefont {M.}~\bibnamefont {{S{\o}rensen}}}, \ and\ \bibinfo {author}
  {\bibfnamefont {H.~F.}\ \bibnamefont {{Song}}},\ }\href {\doibase
  10.1051/0004-6361/201832839} {\bibfield  {journal} {\bibinfo  {journal}
  {\aap}\ }\textbf {\bibinfo {volume} {616}},\ \bibinfo {eid} {A28} (\bibinfo
  {year} {2018})},\ \Eprint {http://arxiv.org/abs/1802.05738} {arXiv:1802.05738
  [astro-ph.SR]} \BibitemShut {NoStop}%
\bibitem [{\citenamefont {{Fishbach}}\ \emph {et~al.}(2017)\citenamefont
  {{Fishbach}}, \citenamefont {{Holz}},\ and\ \citenamefont
  {{Farr}}}]{2017ApJ...840L..24F}%
  \BibitemOpen
  \bibfield  {author} {\bibinfo {author} {\bibfnamefont {M.}~\bibnamefont
  {{Fishbach}}}, \bibinfo {author} {\bibfnamefont {D.~E.}\ \bibnamefont
  {{Holz}}}, \ and\ \bibinfo {author} {\bibfnamefont {B.}~\bibnamefont
  {{Farr}}},\ }\href {\doibase 10.3847/2041-8213/aa7045} {\bibfield  {journal}
  {\bibinfo  {journal} {\apjl}\ }\textbf {\bibinfo {volume} {840}},\ \bibinfo
  {eid} {L24} (\bibinfo {year} {2017})},\ \Eprint
  {http://arxiv.org/abs/1703.06869} {arXiv:1703.06869 [astro-ph.HE]}
  \BibitemShut {NoStop}%
\bibitem [{\citenamefont {{Vaccaro}}\ \emph {et~al.}(2024)\citenamefont
  {{Vaccaro}}, \citenamefont {{Mapelli}}, \citenamefont {{P{\'e}rigois}},
  \citenamefont {{Barone}}, \citenamefont {{Artale}}, \citenamefont
  {{Dall'Amico}}, \citenamefont {{Iorio}},\ and\ \citenamefont
  {{Torniamenti}}}]{2024A&A...685A..51V}%
  \BibitemOpen
  \bibfield  {author} {\bibinfo {author} {\bibfnamefont {M.~P.}\ \bibnamefont
  {{Vaccaro}}}, \bibinfo {author} {\bibfnamefont {M.}~\bibnamefont
  {{Mapelli}}}, \bibinfo {author} {\bibfnamefont {C.}~\bibnamefont
  {{P{\'e}rigois}}}, \bibinfo {author} {\bibfnamefont {D.}~\bibnamefont
  {{Barone}}}, \bibinfo {author} {\bibfnamefont {M.~C.}\ \bibnamefont
  {{Artale}}}, \bibinfo {author} {\bibfnamefont {M.}~\bibnamefont
  {{Dall'Amico}}}, \bibinfo {author} {\bibfnamefont {G.}~\bibnamefont
  {{Iorio}}}, \ and\ \bibinfo {author} {\bibfnamefont {S.}~\bibnamefont
  {{Torniamenti}}},\ }\href {\doibase 10.1051/0004-6361/202348509} {\bibfield
  {journal} {\bibinfo  {journal} {\aap}\ }\textbf {\bibinfo {volume} {685}},\
  \bibinfo {eid} {A51} (\bibinfo {year} {2024})},\ \Eprint
  {http://arxiv.org/abs/2311.18548} {arXiv:2311.18548 [astro-ph.HE]}
  \BibitemShut {NoStop}%
\bibitem [{\citenamefont {{Kimball}}\ \emph {et~al.}(2021)\citenamefont
  {{Kimball}}, \citenamefont {{Talbot}}, \citenamefont {{Berry}}, \citenamefont
  {{Zevin}}, \citenamefont {{Thrane}}, \citenamefont {{Kalogera}},
  \citenamefont {{Buscicchio}}, \citenamefont {{Carney}}, \citenamefont
  {{Dent}}, \citenamefont {{Middleton}}, \citenamefont {{Payne}}, \citenamefont
  {{Veitch}},\ and\ \citenamefont {{Williams}}}]{2021ApJ...915L..35K}%
  \BibitemOpen
  \bibfield  {author} {\bibinfo {author} {\bibfnamefont {C.}~\bibnamefont
  {{Kimball}}}, \bibinfo {author} {\bibfnamefont {C.}~\bibnamefont {{Talbot}}},
  \bibinfo {author} {\bibfnamefont {C.~P.~L.}\ \bibnamefont {{Berry}}},
  \bibinfo {author} {\bibfnamefont {M.}~\bibnamefont {{Zevin}}}, \bibinfo
  {author} {\bibfnamefont {E.}~\bibnamefont {{Thrane}}}, \bibinfo {author}
  {\bibfnamefont {V.}~\bibnamefont {{Kalogera}}}, \bibinfo {author}
  {\bibfnamefont {R.}~\bibnamefont {{Buscicchio}}}, \bibinfo {author}
  {\bibfnamefont {M.}~\bibnamefont {{Carney}}}, \bibinfo {author}
  {\bibfnamefont {T.}~\bibnamefont {{Dent}}}, \bibinfo {author} {\bibfnamefont
  {H.}~\bibnamefont {{Middleton}}}, \bibinfo {author} {\bibfnamefont
  {E.}~\bibnamefont {{Payne}}}, \bibinfo {author} {\bibfnamefont
  {J.}~\bibnamefont {{Veitch}}}, \ and\ \bibinfo {author} {\bibfnamefont
  {D.}~\bibnamefont {{Williams}}},\ }\href {\doibase 10.3847/2041-8213/ac0aef}
  {\bibfield  {journal} {\bibinfo  {journal} {\apjl}\ }\textbf {\bibinfo
  {volume} {915}},\ \bibinfo {eid} {L35} (\bibinfo {year} {2021})},\ \Eprint
  {http://arxiv.org/abs/2011.05332} {arXiv:2011.05332 [astro-ph.HE]}
  \BibitemShut {NoStop}%
\bibitem [{\citenamefont {{Wang}}\ \emph {et~al.}(2022)\citenamefont {{Wang}},
  \citenamefont {{Li}}, \citenamefont {{Vink}}, \citenamefont {{Fan}},
  \citenamefont {{Tang}}, \citenamefont {{Qin}},\ and\ \citenamefont
  {{Wei}}}]{2022ApJ...941L..39W}%
  \BibitemOpen
  \bibfield  {author} {\bibinfo {author} {\bibfnamefont {Y.-Z.}\ \bibnamefont
  {{Wang}}}, \bibinfo {author} {\bibfnamefont {Y.-J.}\ \bibnamefont {{Li}}},
  \bibinfo {author} {\bibfnamefont {J.~S.}\ \bibnamefont {{Vink}}}, \bibinfo
  {author} {\bibfnamefont {Y.-Z.}\ \bibnamefont {{Fan}}}, \bibinfo {author}
  {\bibfnamefont {S.-P.}\ \bibnamefont {{Tang}}}, \bibinfo {author}
  {\bibfnamefont {Y.}~\bibnamefont {{Qin}}}, \ and\ \bibinfo {author}
  {\bibfnamefont {D.-M.}\ \bibnamefont {{Wei}}},\ }\href {\doibase
  10.3847/2041-8213/aca89f} {\bibfield  {journal} {\bibinfo  {journal} {\apjl}\
  }\textbf {\bibinfo {volume} {941}},\ \bibinfo {eid} {L39} (\bibinfo {year}
  {2022})},\ \Eprint {http://arxiv.org/abs/2208.11871} {arXiv:2208.11871
  [astro-ph.HE]} \BibitemShut {NoStop}%
\bibitem [{\citenamefont {{Li}}\ \emph
  {et~al.}(2024{\natexlab{a}})\citenamefont {{Li}}, \citenamefont {{Wang}},
  \citenamefont {{Tang}},\ and\ \citenamefont {{Fan}}}]{2024PhRvL.133e1401L}%
  \BibitemOpen
  \bibfield  {author} {\bibinfo {author} {\bibfnamefont {Y.-J.}\ \bibnamefont
  {{Li}}}, \bibinfo {author} {\bibfnamefont {Y.-Z.}\ \bibnamefont {{Wang}}},
  \bibinfo {author} {\bibfnamefont {S.-P.}\ \bibnamefont {{Tang}}}, \ and\
  \bibinfo {author} {\bibfnamefont {Y.-Z.}\ \bibnamefont {{Fan}}},\ }\href
  {\doibase 10.1103/PhysRevLett.133.051401} {\bibfield  {journal} {\bibinfo
  {journal} {\prl}\ }\textbf {\bibinfo {volume} {133}},\ \bibinfo {eid}
  {051401} (\bibinfo {year} {2024}{\natexlab{a}})},\ \Eprint
  {http://arxiv.org/abs/2303.02973} {arXiv:2303.02973 [astro-ph.HE]}
  \BibitemShut {NoStop}%
\bibitem [{\citenamefont {{Pierra}}\ \emph {et~al.}(2024)\citenamefont
  {{Pierra}}, \citenamefont {{Mastrogiovanni}},\ and\ \citenamefont
  {{Perri{\`e}s}}}]{2024A&A...692A..80P}%
  \BibitemOpen
  \bibfield  {author} {\bibinfo {author} {\bibfnamefont {G.}~\bibnamefont
  {{Pierra}}}, \bibinfo {author} {\bibfnamefont {S.}~\bibnamefont
  {{Mastrogiovanni}}}, \ and\ \bibinfo {author} {\bibfnamefont
  {S.}~\bibnamefont {{Perri{\`e}s}}},\ }\href {\doibase
  10.1051/0004-6361/202452545} {\bibfield  {journal} {\bibinfo  {journal}
  {\aap}\ }\textbf {\bibinfo {volume} {692}},\ \bibinfo {eid} {A80} (\bibinfo
  {year} {2024})}\BibitemShut {NoStop}%
\bibitem [{\citenamefont {{Antonini}}\ \emph
  {et~al.}(2025{\natexlab{a}})\citenamefont {{Antonini}}, \citenamefont
  {{Romero-Shaw}},\ and\ \citenamefont {{Callister}}}]{2025PhRvL.134a1401A}%
  \BibitemOpen
  \bibfield  {author} {\bibinfo {author} {\bibfnamefont {F.}~\bibnamefont
  {{Antonini}}}, \bibinfo {author} {\bibfnamefont {I.~M.}\ \bibnamefont
  {{Romero-Shaw}}}, \ and\ \bibinfo {author} {\bibfnamefont {T.}~\bibnamefont
  {{Callister}}},\ }\href {\doibase 10.1103/PhysRevLett.134.011401} {\bibfield
  {journal} {\bibinfo  {journal} {\prl}\ }\textbf {\bibinfo {volume} {134}},\
  \bibinfo {eid} {011401} (\bibinfo {year} {2025}{\natexlab{a}})},\ \Eprint
  {http://arxiv.org/abs/2406.19044} {arXiv:2406.19044 [astro-ph.HE]}
  \BibitemShut {NoStop}%
\bibitem [{\citenamefont {{Li}}\ \emph {et~al.}(2025)\citenamefont {{Li}},
  \citenamefont {{Wang}}, \citenamefont {{Tang}}, \citenamefont {{Chen}},\ and\
  \citenamefont {{Fan}}}]{2025ApJ...987...65L}%
  \BibitemOpen
  \bibfield  {author} {\bibinfo {author} {\bibfnamefont {Y.-J.}\ \bibnamefont
  {{Li}}}, \bibinfo {author} {\bibfnamefont {Y.-Z.}\ \bibnamefont {{Wang}}},
  \bibinfo {author} {\bibfnamefont {S.-P.}\ \bibnamefont {{Tang}}}, \bibinfo
  {author} {\bibfnamefont {T.}~\bibnamefont {{Chen}}}, \ and\ \bibinfo {author}
  {\bibfnamefont {Y.-Z.}\ \bibnamefont {{Fan}}},\ }\href {\doibase
  10.3847/1538-4357/add535} {\bibfield  {journal} {\bibinfo  {journal} {\apj}\
  }\textbf {\bibinfo {volume} {987}},\ \bibinfo {eid} {65} (\bibinfo {year}
  {2025})},\ \Eprint {http://arxiv.org/abs/2501.09495} {arXiv:2501.09495
  [astro-ph.HE]} \BibitemShut {NoStop}%
\bibitem [{\citenamefont {{Graham}}\ \emph {et~al.}(2020)\citenamefont
  {{Graham}}, \citenamefont {{Ford}}, \citenamefont {{McKernan}}, \citenamefont
  {{Ross}}, \citenamefont {{Stern}}, \citenamefont {{Burdge}}, \citenamefont
  {{Coughlin}}, \citenamefont {{Djorgovski}}, \citenamefont {{Drake}},
  \citenamefont {{Duev}}, \citenamefont {{Kasliwal}}, \citenamefont
  {{Mahabal}}, \citenamefont {{van Velzen}}, \citenamefont {{Belecki}},
  \citenamefont {{Bellm}}, \citenamefont {{Burruss}}, \citenamefont {{Cenko}},
  \citenamefont {{Cunningham}}, \citenamefont {{Helou}}, \citenamefont
  {{Kulkarni}}, \citenamefont {{Masci}}, \citenamefont {{Prince}},
  \citenamefont {{Reiley}}, \citenamefont {{Rodriguez}}, \citenamefont
  {{Rusholme}}, \citenamefont {{Smith}},\ and\ \citenamefont
  {{Soumagnac}}}]{2020PhRvL.124y1102G}%
  \BibitemOpen
  \bibfield  {author} {\bibinfo {author} {\bibfnamefont {M.~J.}\ \bibnamefont
  {{Graham}}}, \bibinfo {author} {\bibfnamefont {K.~E.~S.}\ \bibnamefont
  {{Ford}}}, \bibinfo {author} {\bibfnamefont {B.}~\bibnamefont {{McKernan}}},
  \bibinfo {author} {\bibfnamefont {N.~P.}\ \bibnamefont {{Ross}}}, \bibinfo
  {author} {\bibfnamefont {D.}~\bibnamefont {{Stern}}}, \bibinfo {author}
  {\bibfnamefont {K.}~\bibnamefont {{Burdge}}}, \bibinfo {author}
  {\bibfnamefont {M.}~\bibnamefont {{Coughlin}}}, \bibinfo {author}
  {\bibfnamefont {S.~G.}\ \bibnamefont {{Djorgovski}}}, \bibinfo {author}
  {\bibfnamefont {A.~J.}\ \bibnamefont {{Drake}}}, \bibinfo {author}
  {\bibfnamefont {D.}~\bibnamefont {{Duev}}}, \bibinfo {author} {\bibfnamefont
  {M.}~\bibnamefont {{Kasliwal}}}, \bibinfo {author} {\bibfnamefont {A.~A.}\
  \bibnamefont {{Mahabal}}}, \bibinfo {author} {\bibfnamefont {S.}~\bibnamefont
  {{van Velzen}}}, \bibinfo {author} {\bibfnamefont {J.}~\bibnamefont
  {{Belecki}}}, \bibinfo {author} {\bibfnamefont {E.~C.}\ \bibnamefont
  {{Bellm}}}, \bibinfo {author} {\bibfnamefont {R.}~\bibnamefont {{Burruss}}},
  \bibinfo {author} {\bibfnamefont {S.~B.}\ \bibnamefont {{Cenko}}}, \bibinfo
  {author} {\bibfnamefont {V.}~\bibnamefont {{Cunningham}}}, \bibinfo {author}
  {\bibfnamefont {G.}~\bibnamefont {{Helou}}}, \bibinfo {author} {\bibfnamefont
  {S.~R.}\ \bibnamefont {{Kulkarni}}}, \bibinfo {author} {\bibfnamefont
  {F.~J.}\ \bibnamefont {{Masci}}}, \bibinfo {author} {\bibfnamefont
  {T.}~\bibnamefont {{Prince}}}, \bibinfo {author} {\bibfnamefont
  {D.}~\bibnamefont {{Reiley}}}, \bibinfo {author} {\bibfnamefont
  {H.}~\bibnamefont {{Rodriguez}}}, \bibinfo {author} {\bibfnamefont
  {B.}~\bibnamefont {{Rusholme}}}, \bibinfo {author} {\bibfnamefont {R.~M.}\
  \bibnamefont {{Smith}}}, \ and\ \bibinfo {author} {\bibfnamefont {M.~T.}\
  \bibnamefont {{Soumagnac}}},\ }\href {\doibase
  10.1103/PhysRevLett.124.251102} {\bibfield  {journal} {\bibinfo  {journal}
  {\prl}\ }\textbf {\bibinfo {volume} {124}},\ \bibinfo {eid} {251102}
  (\bibinfo {year} {2020})},\ \Eprint {http://arxiv.org/abs/2006.14122}
  {arXiv:2006.14122 [astro-ph.HE]} \BibitemShut {NoStop}%
\bibitem [{\citenamefont {{Li}}\ \emph {et~al.}(2026)\citenamefont {{Li}},
  \citenamefont {{Tang}}, \citenamefont {{Xue}},\ and\ \citenamefont
  {{Fan}}}]{2026ApJ...999..127L}%
  \BibitemOpen
  \bibfield  {author} {\bibinfo {author} {\bibfnamefont {Y.-J.}\ \bibnamefont
  {{Li}}}, \bibinfo {author} {\bibfnamefont {S.-P.}\ \bibnamefont {{Tang}}},
  \bibinfo {author} {\bibfnamefont {L.-Q.}\ \bibnamefont {{Xue}}}, \ and\
  \bibinfo {author} {\bibfnamefont {Y.-Z.}\ \bibnamefont {{Fan}}},\ }\href
  {\doibase 10.3847/1538-4357/ae4102} {\bibfield  {journal} {\bibinfo
  {journal} {\apj}\ }\textbf {\bibinfo {volume} {999}},\ \bibinfo {eid} {127}
  (\bibinfo {year} {2026})},\ \Eprint {http://arxiv.org/abs/2507.17551}
  {arXiv:2507.17551 [astro-ph.HE]} \BibitemShut {NoStop}%
\bibitem [{\citenamefont {{Abbott}}\ \emph {et~al.}(2020)\citenamefont
  {{Abbott}}, \citenamefont {{Abbott}}, \citenamefont {{Abraham}},
  \citenamefont {{Acernese}} \emph {et~al.}}]{2020ApJ...900L..13A}%
  \BibitemOpen
  \bibfield  {author} {\bibinfo {author} {\bibfnamefont {R.}~\bibnamefont
  {{Abbott}}}, \bibinfo {author} {\bibfnamefont {T.~D.}\ \bibnamefont
  {{Abbott}}}, \bibinfo {author} {\bibfnamefont {S.}~\bibnamefont {{Abraham}}},
  \bibinfo {author} {\bibfnamefont {F.}~\bibnamefont {{Acernese}}},  \emph
  {et~al.},\ }\href {\doibase 10.3847/2041-8213/aba493} {\bibfield  {journal}
  {\bibinfo  {journal} {\apjl}\ }\textbf {\bibinfo {volume} {900}},\ \bibinfo
  {eid} {L13} (\bibinfo {year} {2020})},\ \Eprint
  {http://arxiv.org/abs/2009.01190} {arXiv:2009.01190 [astro-ph.HE]}
  \BibitemShut {NoStop}%
\bibitem [{\citenamefont {{Gamba}}\ \emph {et~al.}(2023)\citenamefont
  {{Gamba}}, \citenamefont {{Breschi}}, \citenamefont {{Carullo}},
  \citenamefont {{Albanesi}}, \citenamefont {{Rettegno}}, \citenamefont
  {{Bernuzzi}},\ and\ \citenamefont {{Nagar}}}]{2023NatAs...7...11G}%
  \BibitemOpen
  \bibfield  {author} {\bibinfo {author} {\bibfnamefont {R.}~\bibnamefont
  {{Gamba}}}, \bibinfo {author} {\bibfnamefont {M.}~\bibnamefont {{Breschi}}},
  \bibinfo {author} {\bibfnamefont {G.}~\bibnamefont {{Carullo}}}, \bibinfo
  {author} {\bibfnamefont {S.}~\bibnamefont {{Albanesi}}}, \bibinfo {author}
  {\bibfnamefont {P.}~\bibnamefont {{Rettegno}}}, \bibinfo {author}
  {\bibfnamefont {S.}~\bibnamefont {{Bernuzzi}}}, \ and\ \bibinfo {author}
  {\bibfnamefont {A.}~\bibnamefont {{Nagar}}},\ }\href {\doibase
  10.1038/s41550-022-01813-w} {\bibfield  {journal} {\bibinfo  {journal}
  {Nature Astronomy}\ }\textbf {\bibinfo {volume} {7}},\ \bibinfo {pages} {11}
  (\bibinfo {year} {2023})},\ \Eprint {http://arxiv.org/abs/2106.05575}
  {arXiv:2106.05575 [gr-qc]} \BibitemShut {NoStop}%
\bibitem [{\citenamefont {{Abac}}\ \emph {et~al.}(2025)\citenamefont {{Abac}},
  \citenamefont {{Abouelfettouh}}, \citenamefont {{Acernese}}, \citenamefont
  {{Ackley}}, \citenamefont {{Adamcewicz}} \emph
  {et~al.}}]{2025ApJ...993L..25A}%
  \BibitemOpen
  \bibfield  {author} {\bibinfo {author} {\bibfnamefont {A.~G.}\ \bibnamefont
  {{Abac}}}, \bibinfo {author} {\bibfnamefont {I.}~\bibnamefont
  {{Abouelfettouh}}}, \bibinfo {author} {\bibfnamefont {F.}~\bibnamefont
  {{Acernese}}}, \bibinfo {author} {\bibfnamefont {K.}~\bibnamefont
  {{Ackley}}}, \bibinfo {author} {\bibfnamefont {C.}~\bibnamefont
  {{Adamcewicz}}},  \emph {et~al.},\ }\href {\doibase 10.3847/2041-8213/ae0c9c}
  {\bibfield  {journal} {\bibinfo  {journal} {\apjl}\ }\textbf {\bibinfo
  {volume} {993}},\ \bibinfo {eid} {L25} (\bibinfo {year} {2025})},\ \Eprint
  {http://arxiv.org/abs/2507.08219} {arXiv:2507.08219 [astro-ph.HE]}
  \BibitemShut {NoStop}%
\bibitem [{\citenamefont {{K{\i}ro{\u{g}}lu}}\ \emph
  {et~al.}(2025)\citenamefont {{K{\i}ro{\u{g}}lu}}, \citenamefont {{Kremer}},\
  and\ \citenamefont {{Rasio}}}]{2025ApJ...994L..37K}%
  \BibitemOpen
  \bibfield  {author} {\bibinfo {author} {\bibfnamefont {F.}~\bibnamefont
  {{K{\i}ro{\u{g}}lu}}}, \bibinfo {author} {\bibfnamefont {K.}~\bibnamefont
  {{Kremer}}}, \ and\ \bibinfo {author} {\bibfnamefont {F.~A.}\ \bibnamefont
  {{Rasio}}},\ }\href {\doibase 10.3847/2041-8213/ae1eeb} {\bibfield  {journal}
  {\bibinfo  {journal} {\apjl}\ }\textbf {\bibinfo {volume} {994}},\ \bibinfo
  {eid} {L37} (\bibinfo {year} {2025})},\ \Eprint
  {http://arxiv.org/abs/2509.05415} {arXiv:2509.05415 [astro-ph.HE]}
  \BibitemShut {NoStop}%
\bibitem [{\citenamefont {{Yang}}\ \emph
  {et~al.}(2019{\natexlab{a}})\citenamefont {{Yang}}, \citenamefont {{Bartos}},
  \citenamefont {{Gayathri}}, \citenamefont {{Ford}}, \citenamefont {{Haiman}},
  \citenamefont {{Klimenko}}, \citenamefont {{Kocsis}}, \citenamefont
  {{M{\'a}rka}}, \citenamefont {{M{\'a}rka}}, \citenamefont {{McKernan}},\ and\
  \citenamefont {{O'Shaughnessy}}}]{2019PhRvL.123r1101Y}%
  \BibitemOpen
  \bibfield  {author} {\bibinfo {author} {\bibfnamefont {Y.}~\bibnamefont
  {{Yang}}}, \bibinfo {author} {\bibfnamefont {I.}~\bibnamefont {{Bartos}}},
  \bibinfo {author} {\bibfnamefont {V.}~\bibnamefont {{Gayathri}}}, \bibinfo
  {author} {\bibfnamefont {K.~E.~S.}\ \bibnamefont {{Ford}}}, \bibinfo {author}
  {\bibfnamefont {Z.}~\bibnamefont {{Haiman}}}, \bibinfo {author}
  {\bibfnamefont {S.}~\bibnamefont {{Klimenko}}}, \bibinfo {author}
  {\bibfnamefont {B.}~\bibnamefont {{Kocsis}}}, \bibinfo {author}
  {\bibfnamefont {S.}~\bibnamefont {{M{\'a}rka}}}, \bibinfo {author}
  {\bibfnamefont {Z.}~\bibnamefont {{M{\'a}rka}}}, \bibinfo {author}
  {\bibfnamefont {B.}~\bibnamefont {{McKernan}}}, \ and\ \bibinfo {author}
  {\bibfnamefont {R.}~\bibnamefont {{O'Shaughnessy}}},\ }\href {\doibase
  10.1103/PhysRevLett.123.181101} {\bibfield  {journal} {\bibinfo  {journal}
  {\prl}\ }\textbf {\bibinfo {volume} {123}},\ \bibinfo {eid} {181101}
  (\bibinfo {year} {2019}{\natexlab{a}})},\ \Eprint
  {http://arxiv.org/abs/1906.09281} {arXiv:1906.09281 [astro-ph.HE]}
  \BibitemShut {NoStop}%
\bibitem [{\citenamefont {{Wang}}\ \emph
  {et~al.}(2021{\natexlab{a}})\citenamefont {{Wang}}, \citenamefont
  {{McKernan}}, \citenamefont {{Ford}}, \citenamefont {{Perna}}, \citenamefont
  {{Leigh}},\ and\ \citenamefont {{Mac Low}}}]{2021ApJ...923L..23W}%
  \BibitemOpen
  \bibfield  {author} {\bibinfo {author} {\bibfnamefont {Y.-H.}\ \bibnamefont
  {{Wang}}}, \bibinfo {author} {\bibfnamefont {B.}~\bibnamefont {{McKernan}}},
  \bibinfo {author} {\bibfnamefont {S.}~\bibnamefont {{Ford}}}, \bibinfo
  {author} {\bibfnamefont {R.}~\bibnamefont {{Perna}}}, \bibinfo {author}
  {\bibfnamefont {N.~W.~C.}\ \bibnamefont {{Leigh}}}, \ and\ \bibinfo {author}
  {\bibfnamefont {M.-M.}\ \bibnamefont {{Mac Low}}},\ }\href {\doibase
  10.3847/2041-8213/ac400a} {\bibfield  {journal} {\bibinfo  {journal} {\apjl}\
  }\textbf {\bibinfo {volume} {923}},\ \bibinfo {eid} {L23} (\bibinfo {year}
  {2021}{\natexlab{a}})},\ \Eprint {http://arxiv.org/abs/2110.03698}
  {arXiv:2110.03698 [astro-ph.HE]} \BibitemShut {NoStop}%
\bibitem [{\citenamefont {{Samsing}}\ \emph {et~al.}(2022)\citenamefont
  {{Samsing}}, \citenamefont {{Bartos}}, \citenamefont {{D'Orazio}},
  \citenamefont {{Haiman}}, \citenamefont {{Kocsis}}, \citenamefont {{Leigh}},
  \citenamefont {{Liu}}, \citenamefont {{Pessah}},\ and\ \citenamefont
  {{Tagawa}}}]{2022Natur.603..237S}%
  \BibitemOpen
  \bibfield  {author} {\bibinfo {author} {\bibfnamefont {J.}~\bibnamefont
  {{Samsing}}}, \bibinfo {author} {\bibfnamefont {I.}~\bibnamefont {{Bartos}}},
  \bibinfo {author} {\bibfnamefont {D.~J.}\ \bibnamefont {{D'Orazio}}},
  \bibinfo {author} {\bibfnamefont {Z.}~\bibnamefont {{Haiman}}}, \bibinfo
  {author} {\bibfnamefont {B.}~\bibnamefont {{Kocsis}}}, \bibinfo {author}
  {\bibfnamefont {N.~W.~C.}\ \bibnamefont {{Leigh}}}, \bibinfo {author}
  {\bibfnamefont {B.}~\bibnamefont {{Liu}}}, \bibinfo {author} {\bibfnamefont
  {M.~E.}\ \bibnamefont {{Pessah}}}, \ and\ \bibinfo {author} {\bibfnamefont
  {H.}~\bibnamefont {{Tagawa}}},\ }\href {\doibase 10.1038/s41586-021-04333-1}
  {\bibfield  {journal} {\bibinfo  {journal} {\nat}\ }\textbf {\bibinfo
  {volume} {603}},\ \bibinfo {pages} {237} (\bibinfo {year} {2022})},\ \Eprint
  {http://arxiv.org/abs/2010.09765} {arXiv:2010.09765 [astro-ph.HE]}
  \BibitemShut {NoStop}%
\bibitem [{\citenamefont {{McKernan}}\ and\ \citenamefont
  {{Ford}}(2024)}]{2024MNRAS.531.3479M}%
  \BibitemOpen
  \bibfield  {author} {\bibinfo {author} {\bibfnamefont {B.}~\bibnamefont
  {{McKernan}}}\ and\ \bibinfo {author} {\bibfnamefont {K.~E.~S.}\ \bibnamefont
  {{Ford}}},\ }\href {\doibase 10.1093/mnras/stae1351} {\bibfield  {journal}
  {\bibinfo  {journal} {\mnras}\ }\textbf {\bibinfo {volume} {531}},\ \bibinfo
  {pages} {3479} (\bibinfo {year} {2024})},\ \Eprint
  {http://arxiv.org/abs/2309.15213} {arXiv:2309.15213 [astro-ph.HE]}
  \BibitemShut {NoStop}%
\bibitem [{\citenamefont {{Wang}}\ \emph
  {et~al.}(2021{\natexlab{b}})\citenamefont {{Wang}}, \citenamefont {{Fan}},
  \citenamefont {{Tang}}, \citenamefont {{Qin}},\ and\ \citenamefont
  {{Wei}}}]{2021arXiv211010838W}%
  \BibitemOpen
  \bibfield  {author} {\bibinfo {author} {\bibfnamefont {Y.-Z.}\ \bibnamefont
  {{Wang}}}, \bibinfo {author} {\bibfnamefont {Y.-Z.}\ \bibnamefont {{Fan}}},
  \bibinfo {author} {\bibfnamefont {S.-P.}\ \bibnamefont {{Tang}}}, \bibinfo
  {author} {\bibfnamefont {Y.}~\bibnamefont {{Qin}}}, \ and\ \bibinfo {author}
  {\bibfnamefont {D.-M.}\ \bibnamefont {{Wei}}},\ }\href {\doibase
  10.48550/arXiv.2110.10838} {\bibfield  {journal} {\bibinfo  {journal} {arXiv
  e-prints}\ ,\ \bibinfo {eid} {arXiv:2110.10838}} (\bibinfo {year}
  {2021}{\natexlab{b}})},\ \Eprint {http://arxiv.org/abs/2110.10838}
  {arXiv:2110.10838 [astro-ph.HE]} \BibitemShut {NoStop}%
\bibitem [{\citenamefont {{McKernan}}\ \emph {et~al.}(2025)\citenamefont
  {{McKernan}}, \citenamefont {{Ford}}, \citenamefont {{Cook}}, \citenamefont
  {{Delfavero}}, \citenamefont {{McPike}}, \citenamefont {{Nathaniel}},
  \citenamefont {{Postiglione}}, \citenamefont {{Ray}},\ and\ \citenamefont
  {{O'Shaughnessy}}}]{2025ApJ...990..217M}%
  \BibitemOpen
  \bibfield  {author} {\bibinfo {author} {\bibfnamefont {B.}~\bibnamefont
  {{McKernan}}}, \bibinfo {author} {\bibfnamefont {K.~E.~S.}\ \bibnamefont
  {{Ford}}}, \bibinfo {author} {\bibfnamefont {H.~E.}\ \bibnamefont {{Cook}}},
  \bibinfo {author} {\bibfnamefont {V.}~\bibnamefont {{Delfavero}}}, \bibinfo
  {author} {\bibfnamefont {E.}~\bibnamefont {{McPike}}}, \bibinfo {author}
  {\bibfnamefont {K.}~\bibnamefont {{Nathaniel}}}, \bibinfo {author}
  {\bibfnamefont {J.}~\bibnamefont {{Postiglione}}}, \bibinfo {author}
  {\bibfnamefont {S.}~\bibnamefont {{Ray}}}, \ and\ \bibinfo {author}
  {\bibfnamefont {R.}~\bibnamefont {{O'Shaughnessy}}},\ }\href {\doibase
  10.3847/1538-4357/adf114} {\bibfield  {journal} {\bibinfo  {journal} {\apj}\
  }\textbf {\bibinfo {volume} {990}},\ \bibinfo {eid} {217} (\bibinfo {year}
  {2025})},\ \Eprint {http://arxiv.org/abs/2410.16515} {arXiv:2410.16515
  [astro-ph.HE]} \BibitemShut {NoStop}%
\bibitem [{\citenamefont {{Zevin}}\ \emph {et~al.}(2021)\citenamefont
  {{Zevin}}, \citenamefont {{Bavera}}, \citenamefont {{Berry}}, \citenamefont
  {{Kalogera}}, \citenamefont {{Fragos}}, \citenamefont {{Marchant}},
  \citenamefont {{Rodriguez}}, \citenamefont {{Antonini}}, \citenamefont
  {{Holz}},\ and\ \citenamefont {{Pankow}}}]{2021ApJ...910..152Z}%
  \BibitemOpen
  \bibfield  {author} {\bibinfo {author} {\bibfnamefont {M.}~\bibnamefont
  {{Zevin}}}, \bibinfo {author} {\bibfnamefont {S.~S.}\ \bibnamefont
  {{Bavera}}}, \bibinfo {author} {\bibfnamefont {C.~P.~L.}\ \bibnamefont
  {{Berry}}}, \bibinfo {author} {\bibfnamefont {V.}~\bibnamefont {{Kalogera}}},
  \bibinfo {author} {\bibfnamefont {T.}~\bibnamefont {{Fragos}}}, \bibinfo
  {author} {\bibfnamefont {P.}~\bibnamefont {{Marchant}}}, \bibinfo {author}
  {\bibfnamefont {C.~L.}\ \bibnamefont {{Rodriguez}}}, \bibinfo {author}
  {\bibfnamefont {F.}~\bibnamefont {{Antonini}}}, \bibinfo {author}
  {\bibfnamefont {D.~E.}\ \bibnamefont {{Holz}}}, \ and\ \bibinfo {author}
  {\bibfnamefont {C.}~\bibnamefont {{Pankow}}},\ }\href {\doibase
  10.3847/1538-4357/abe40e} {\bibfield  {journal} {\bibinfo  {journal} {\apj}\
  ,\ \bibinfo {eid} {152}} (\bibinfo {year} {2021})},\ \Eprint
  {http://arxiv.org/abs/2011.10057} {arXiv:2011.10057 [astro-ph.HE]}
  \BibitemShut {NoStop}%
\bibitem [{\citenamefont {{The LIGO Scientific Collaboration}}\ \emph
  {et~al.}(2025{\natexlab{b}})\citenamefont {{The LIGO Scientific
  Collaboration}}, \citenamefont {{the Virgo Collaboration}},\ and\
  \citenamefont {{the KAGRA Collaboration}}}]{2025arXiv250818083T}%
  \BibitemOpen
  \bibfield  {author} {\bibinfo {author} {\bibnamefont {{The LIGO Scientific
  Collaboration}}}, \bibinfo {author} {\bibnamefont {{the Virgo
  Collaboration}}}, \ and\ \bibinfo {author} {\bibnamefont {{the KAGRA
  Collaboration}}},\ }\href {\doibase 10.48550/arXiv.2508.18083} {\bibfield
  {journal} {\bibinfo  {journal} {arXiv e-prints}\ ,\ \bibinfo {eid}
  {arXiv:2508.18083}} (\bibinfo {year} {2025}{\natexlab{b}})},\ \Eprint
  {http://arxiv.org/abs/2508.18083} {arXiv:2508.18083 [astro-ph.HE]}
  \BibitemShut {NoStop}%
\bibitem [{\citenamefont {{Mandel}}\ \emph {et~al.}(2019)\citenamefont
  {{Mandel}}, \citenamefont {{Farr}},\ and\ \citenamefont
  {{Gair}}}]{2019MNRAS.486.1086M}%
  \BibitemOpen
  \bibfield  {author} {\bibinfo {author} {\bibfnamefont {I.}~\bibnamefont
  {{Mandel}}}, \bibinfo {author} {\bibfnamefont {W.~M.}\ \bibnamefont
  {{Farr}}}, \ and\ \bibinfo {author} {\bibfnamefont {J.~R.}\ \bibnamefont
  {{Gair}}},\ }\href {\doibase 10.1093/mnras/stz896} {\bibfield  {journal}
  {\bibinfo  {journal} {\mnras}\ ,\ \bibinfo {pages} {1086}} (\bibinfo {year}
  {2019})},\ \Eprint {http://arxiv.org/abs/1809.02063} {arXiv:1809.02063
  [physics.data-an]} \BibitemShut {NoStop}%
\bibitem [{sup()}]{supp}%
  \BibitemOpen
  \href@noop {} {}\bibinfo {note} {{See Supplemental Material at [URL] for some
  supplemental data analyses, discussion, and figures, which includes Refs.
  \citep{2023PhRvX..13a1048A,2025PhRvD.112j2001E,2016ascl.soft06005B,2022ApJ...924..101E,2021ApJ...913L...7A,2021ApJ...917...33L,2024MNRAS.527..298T,2013PhRvD..87b4035B,2017PhRvD..96l4041W,2019MNRAS.484.4216R,2021ApJ...922L...5C,2024PhRvD.109j4036M,2023MNRAS.526.3495T}.}}\BibitemShut
  {Stop}%
\bibitem [{\citenamefont {{Abbott}}\ \emph
  {et~al.}(2023{\natexlab{b}})\citenamefont {{Abbott}}, \citenamefont
  {{Abbott}}, \citenamefont {{Acernese}}, \citenamefont {{Ackley}} \emph
  {et~al.}}]{2023PhRvX..13a1048A}%
  \BibitemOpen
  \bibfield  {author} {\bibinfo {author} {\bibfnamefont {R.}~\bibnamefont
  {{Abbott}}}, \bibinfo {author} {\bibfnamefont {T.~D.}\ \bibnamefont
  {{Abbott}}}, \bibinfo {author} {\bibfnamefont {F.}~\bibnamefont
  {{Acernese}}}, \bibinfo {author} {\bibfnamefont {K.}~\bibnamefont
  {{Ackley}}},  \emph {et~al.},\ }\href {\doibase 10.1103/PhysRevX.13.011048}
  {\bibfield  {journal} {\bibinfo  {journal} {Physical Review X}\ }\textbf
  {\bibinfo {volume} {13}},\ \bibinfo {eid} {011048} (\bibinfo {year}
  {2023}{\natexlab{b}})},\ \Eprint {http://arxiv.org/abs/2111.03634}
  {arXiv:2111.03634 [astro-ph.HE]} \BibitemShut {NoStop}%
\bibitem [{\citenamefont {{Essick}}\ \emph {et~al.}(2025)\citenamefont
  {{Essick}}, \citenamefont {{Coughlin}}, \citenamefont {{Zevin}},
  \citenamefont {{Chatterjee}}, \citenamefont {{Clarke}}, \citenamefont
  {{Colloms}}, \citenamefont {{Mali}}, \citenamefont {{Miller}}, \citenamefont
  {{Steinle}}, \citenamefont {{Baral}}, \citenamefont {{Baylor}}, \citenamefont
  {{Cabourn Davies}}, \citenamefont {{Dent}}, \citenamefont {{Joshi}},
  \citenamefont {{Kumar}}, \citenamefont {{Messick}}, \citenamefont {{Mishra}},
  \citenamefont {{Ouzriat}}, \citenamefont {{Phukon}}, \citenamefont
  {{Piccari}}, \citenamefont {{Pillas}}, \citenamefont {{Trevor}},
  \citenamefont {{Callister}},\ and\ \citenamefont
  {{Fishbach}}}]{2025PhRvD.112j2001E}%
  \BibitemOpen
  \bibfield  {author} {\bibinfo {author} {\bibfnamefont {R.}~\bibnamefont
  {{Essick}}}, \bibinfo {author} {\bibfnamefont {M.~W.}\ \bibnamefont
  {{Coughlin}}}, \bibinfo {author} {\bibfnamefont {M.}~\bibnamefont {{Zevin}}},
  \bibinfo {author} {\bibfnamefont {D.}~\bibnamefont {{Chatterjee}}}, \bibinfo
  {author} {\bibfnamefont {T.~A.}\ \bibnamefont {{Clarke}}}, \bibinfo {author}
  {\bibfnamefont {S.}~\bibnamefont {{Colloms}}}, \bibinfo {author}
  {\bibfnamefont {U.}~\bibnamefont {{Mali}}}, \bibinfo {author} {\bibfnamefont
  {S.}~\bibnamefont {{Miller}}}, \bibinfo {author} {\bibfnamefont
  {N.}~\bibnamefont {{Steinle}}}, \bibinfo {author} {\bibfnamefont
  {P.}~\bibnamefont {{Baral}}}, \bibinfo {author} {\bibfnamefont {A.~C.}\
  \bibnamefont {{Baylor}}}, \bibinfo {author} {\bibfnamefont {G.}~\bibnamefont
  {{Cabourn Davies}}}, \bibinfo {author} {\bibfnamefont {T.}~\bibnamefont
  {{Dent}}}, \bibinfo {author} {\bibfnamefont {P.}~\bibnamefont {{Joshi}}},
  \bibinfo {author} {\bibfnamefont {P.}~\bibnamefont {{Kumar}}}, \bibinfo
  {author} {\bibfnamefont {C.}~\bibnamefont {{Messick}}}, \bibinfo {author}
  {\bibfnamefont {T.}~\bibnamefont {{Mishra}}}, \bibinfo {author}
  {\bibfnamefont {A.}~\bibnamefont {{Ouzriat}}}, \bibinfo {author}
  {\bibfnamefont {K.~S.}\ \bibnamefont {{Phukon}}}, \bibinfo {author}
  {\bibfnamefont {L.}~\bibnamefont {{Piccari}}}, \bibinfo {author}
  {\bibfnamefont {M.}~\bibnamefont {{Pillas}}}, \bibinfo {author}
  {\bibfnamefont {M.}~\bibnamefont {{Trevor}}}, \bibinfo {author}
  {\bibfnamefont {T.~A.}\ \bibnamefont {{Callister}}}, \ and\ \bibinfo {author}
  {\bibfnamefont {M.}~\bibnamefont {{Fishbach}}},\ }\href {\doibase
  10.1103/44x3-hv3y} {\bibfield  {journal} {\bibinfo  {journal} {\prd}\
  }\textbf {\bibinfo {volume} {112}},\ \bibinfo {eid} {102001} (\bibinfo {year}
  {2025})},\ \Eprint {http://arxiv.org/abs/2508.10638} {arXiv:2508.10638
  [gr-qc]} \BibitemShut {NoStop}%
\bibitem [{\citenamefont {{Buchner}}(2016)}]{2016ascl.soft06005B}%
  \BibitemOpen
  \bibfield  {author} {\bibinfo {author} {\bibfnamefont {J.}~\bibnamefont
  {{Buchner}}},\ }\href@noop {} {\enquote {\bibinfo {title} {{PyMultiNest:
  Python interface for MultiNest}},}\ }\bibinfo {howpublished} {Astrophysics
  Source Code Library, record ascl:1606.005} (\bibinfo {year} {2016}),\ \Eprint
  {http://arxiv.org/abs/1606.005} {ascl:1606.005} \BibitemShut {NoStop}%
\bibitem [{\citenamefont {{Edelman}}\ \emph {et~al.}(2022)\citenamefont
  {{Edelman}}, \citenamefont {{Doctor}}, \citenamefont {{Godfrey}},\ and\
  \citenamefont {{Farr}}}]{2022ApJ...924..101E}%
  \BibitemOpen
  \bibfield  {author} {\bibinfo {author} {\bibfnamefont {B.}~\bibnamefont
  {{Edelman}}}, \bibinfo {author} {\bibfnamefont {Z.}~\bibnamefont {{Doctor}}},
  \bibinfo {author} {\bibfnamefont {J.}~\bibnamefont {{Godfrey}}}, \ and\
  \bibinfo {author} {\bibfnamefont {B.}~\bibnamefont {{Farr}}},\ }\href
  {\doibase 10.3847/1538-4357/ac3667} {\bibfield  {journal} {\bibinfo
  {journal} {\apj}\ }\textbf {\bibinfo {volume} {924}},\ \bibinfo {eid} {101}
  (\bibinfo {year} {2022})},\ \Eprint {http://arxiv.org/abs/2109.06137}
  {arXiv:2109.06137 [astro-ph.HE]} \BibitemShut {NoStop}%
\bibitem [{\citenamefont {{Abbott}}\ \emph {et~al.}(2021)\citenamefont
  {{Abbott}}, \citenamefont {{Abbott}}, \citenamefont {{Abraham}} \emph
  {et~al.}}]{2021ApJ...913L...7A}%
  \BibitemOpen
  \bibfield  {author} {\bibinfo {author} {\bibfnamefont {R.}~\bibnamefont
  {{Abbott}}}, \bibinfo {author} {\bibfnamefont {T.~D.}\ \bibnamefont
  {{Abbott}}}, \bibinfo {author} {\bibfnamefont {S.}~\bibnamefont {{Abraham}}},
   \emph {et~al.},\ }\href {\doibase 10.3847/2041-8213/abe949} {\bibfield
  {journal} {\bibinfo  {journal} {\apjl}\ }\textbf {\bibinfo {volume} {913}},\
  \bibinfo {eid} {L7} (\bibinfo {year} {2021})},\ \Eprint
  {http://arxiv.org/abs/2010.14533} {arXiv:2010.14533 [astro-ph.HE]}
  \BibitemShut {NoStop}%
\bibitem [{\citenamefont {{Li}}\ \emph {et~al.}(2021)\citenamefont {{Li}},
  \citenamefont {{Wang}}, \citenamefont {{Han}}, \citenamefont {{Tang}},
  \citenamefont {{Yuan}}, \citenamefont {{Fan}},\ and\ \citenamefont
  {{Wei}}}]{2021ApJ...917...33L}%
  \BibitemOpen
  \bibfield  {author} {\bibinfo {author} {\bibfnamefont {Y.-J.}\ \bibnamefont
  {{Li}}}, \bibinfo {author} {\bibfnamefont {Y.-Z.}\ \bibnamefont {{Wang}}},
  \bibinfo {author} {\bibfnamefont {M.-Z.}\ \bibnamefont {{Han}}}, \bibinfo
  {author} {\bibfnamefont {S.-P.}\ \bibnamefont {{Tang}}}, \bibinfo {author}
  {\bibfnamefont {Q.}~\bibnamefont {{Yuan}}}, \bibinfo {author} {\bibfnamefont
  {Y.-Z.}\ \bibnamefont {{Fan}}}, \ and\ \bibinfo {author} {\bibfnamefont
  {D.-M.}\ \bibnamefont {{Wei}}},\ }\href {\doibase 10.3847/1538-4357/ac0971}
  {\bibfield  {journal} {\bibinfo  {journal} {\apj}\ }\textbf {\bibinfo
  {volume} {917}},\ \bibinfo {eid} {33} (\bibinfo {year} {2021})},\ \Eprint
  {http://arxiv.org/abs/2104.02969} {arXiv:2104.02969 [astro-ph.HE]}
  \BibitemShut {NoStop}%
\bibitem [{\citenamefont {{Tiwari}}(2024)}]{2024MNRAS.527..298T}%
  \BibitemOpen
  \bibfield  {author} {\bibinfo {author} {\bibfnamefont {V.}~\bibnamefont
  {{Tiwari}}},\ }\href {\doibase 10.1093/mnras/stad3155} {\bibfield  {journal}
  {\bibinfo  {journal} {\mnras}\ }\textbf {\bibinfo {volume} {527}},\ \bibinfo
  {pages} {298} (\bibinfo {year} {2024})},\ \Eprint
  {http://arxiv.org/abs/2304.03498} {arXiv:2304.03498 [astro-ph.HE]}
  \BibitemShut {NoStop}%
\bibitem [{\citenamefont {{Baird}}\ \emph {et~al.}(2013)\citenamefont
  {{Baird}}, \citenamefont {{Fairhurst}}, \citenamefont {{Hannam}},\ and\
  \citenamefont {{Murphy}}}]{2013PhRvD..87b4035B}%
  \BibitemOpen
  \bibfield  {author} {\bibinfo {author} {\bibfnamefont {E.}~\bibnamefont
  {{Baird}}}, \bibinfo {author} {\bibfnamefont {S.}~\bibnamefont
  {{Fairhurst}}}, \bibinfo {author} {\bibfnamefont {M.}~\bibnamefont
  {{Hannam}}}, \ and\ \bibinfo {author} {\bibfnamefont {P.}~\bibnamefont
  {{Murphy}}},\ }\href {\doibase 10.1103/PhysRevD.87.024035} {\bibfield
  {journal} {\bibinfo  {journal} {\prd}\ }\textbf {\bibinfo {volume} {87}},\
  \bibinfo {eid} {024035} (\bibinfo {year} {2013})},\ \Eprint
  {http://arxiv.org/abs/1211.0546} {arXiv:1211.0546 [gr-qc]} \BibitemShut
  {NoStop}%
\bibitem [{\citenamefont {{Williamson}}\ \emph {et~al.}(2017)\citenamefont
  {{Williamson}}, \citenamefont {{Lange}}, \citenamefont {{O'Shaughnessy}},
  \citenamefont {{Clark}}, \citenamefont {{Kumar}}, \citenamefont
  {{Calder{\'o}n Bustillo}},\ and\ \citenamefont
  {{Veitch}}}]{2017PhRvD..96l4041W}%
  \BibitemOpen
  \bibfield  {author} {\bibinfo {author} {\bibfnamefont {A.~R.}\ \bibnamefont
  {{Williamson}}}, \bibinfo {author} {\bibfnamefont {J.}~\bibnamefont
  {{Lange}}}, \bibinfo {author} {\bibfnamefont {R.}~\bibnamefont
  {{O'Shaughnessy}}}, \bibinfo {author} {\bibfnamefont {J.~A.}\ \bibnamefont
  {{Clark}}}, \bibinfo {author} {\bibfnamefont {P.}~\bibnamefont {{Kumar}}},
  \bibinfo {author} {\bibfnamefont {J.}~\bibnamefont {{Calder{\'o}n
  Bustillo}}}, \ and\ \bibinfo {author} {\bibfnamefont {J.}~\bibnamefont
  {{Veitch}}},\ }\href {\doibase 10.1103/PhysRevD.96.124041} {\bibfield
  {journal} {\bibinfo  {journal} {\prd}\ }\textbf {\bibinfo {volume} {96}},\
  \bibinfo {eid} {124041} (\bibinfo {year} {2017})},\ \Eprint
  {http://arxiv.org/abs/1709.03095} {arXiv:1709.03095 [gr-qc]} \BibitemShut
  {NoStop}%
\bibitem [{\citenamefont {{Roulet}}\ and\ \citenamefont
  {{Zaldarriaga}}(2019)}]{2019MNRAS.484.4216R}%
  \BibitemOpen
  \bibfield  {author} {\bibinfo {author} {\bibfnamefont {J.}~\bibnamefont
  {{Roulet}}}\ and\ \bibinfo {author} {\bibfnamefont {M.}~\bibnamefont
  {{Zaldarriaga}}},\ }\href {\doibase 10.1093/mnras/stz226} {\bibfield
  {journal} {\bibinfo  {journal} {\mnras}\ }\textbf {\bibinfo {volume} {484}},\
  \bibinfo {pages} {4216} (\bibinfo {year} {2019})},\ \Eprint
  {http://arxiv.org/abs/1806.10610} {arXiv:1806.10610 [astro-ph.HE]}
  \BibitemShut {NoStop}%
\bibitem [{\citenamefont {{Callister}}\ \emph {et~al.}(2021)\citenamefont
  {{Callister}}, \citenamefont {{Haster}}, \citenamefont {{Ng}}, \citenamefont
  {{Vitale}},\ and\ \citenamefont {{Farr}}}]{2021ApJ...922L...5C}%
  \BibitemOpen
  \bibfield  {author} {\bibinfo {author} {\bibfnamefont {T.~A.}\ \bibnamefont
  {{Callister}}}, \bibinfo {author} {\bibfnamefont {C.-J.}\ \bibnamefont
  {{Haster}}}, \bibinfo {author} {\bibfnamefont {K.~K.~Y.}\ \bibnamefont
  {{Ng}}}, \bibinfo {author} {\bibfnamefont {S.}~\bibnamefont {{Vitale}}}, \
  and\ \bibinfo {author} {\bibfnamefont {W.~M.}\ \bibnamefont {{Farr}}},\
  }\href {\doibase 10.3847/2041-8213/ac2ccc} {\bibfield  {journal} {\bibinfo
  {journal} {\apjl}\ }\textbf {\bibinfo {volume} {922}},\ \bibinfo {eid} {L5}
  (\bibinfo {year} {2021})},\ \Eprint {http://arxiv.org/abs/2106.00521}
  {arXiv:2106.00521 [astro-ph.HE]} \BibitemShut {NoStop}%
\bibitem [{\citenamefont {{Miller}}\ \emph {et~al.}(2024)\citenamefont
  {{Miller}}, \citenamefont {{Ko}}, \citenamefont {{Callister}},\ and\
  \citenamefont {{Chatziioannou}}}]{2024PhRvD.109j4036M}%
  \BibitemOpen
  \bibfield  {author} {\bibinfo {author} {\bibfnamefont {S.~J.}\ \bibnamefont
  {{Miller}}}, \bibinfo {author} {\bibfnamefont {Z.}~\bibnamefont {{Ko}}},
  \bibinfo {author} {\bibfnamefont {T.}~\bibnamefont {{Callister}}}, \ and\
  \bibinfo {author} {\bibfnamefont {K.}~\bibnamefont {{Chatziioannou}}},\
  }\href {\doibase 10.1103/PhysRevD.109.104036} {\bibfield  {journal} {\bibinfo
   {journal} {\prd}\ }\textbf {\bibinfo {volume} {109}},\ \bibinfo {eid}
  {104036} (\bibinfo {year} {2024})},\ \Eprint
  {http://arxiv.org/abs/2401.05613} {arXiv:2401.05613 [gr-qc]} \BibitemShut
  {NoStop}%
\bibitem [{\citenamefont {{Talbot}}\ and\ \citenamefont
  {{Golomb}}(2023)}]{2023MNRAS.526.3495T}%
  \BibitemOpen
  \bibfield  {author} {\bibinfo {author} {\bibfnamefont {C.}~\bibnamefont
  {{Talbot}}}\ and\ \bibinfo {author} {\bibfnamefont {J.}~\bibnamefont
  {{Golomb}}},\ }\href {\doibase 10.1093/mnras/stad2968} {\bibfield  {journal}
  {\bibinfo  {journal} {\mnras}\ }\textbf {\bibinfo {volume} {526}},\ \bibinfo
  {pages} {3495} (\bibinfo {year} {2023})},\ \Eprint
  {http://arxiv.org/abs/2304.06138} {arXiv:2304.06138 [astro-ph.IM]}
  \BibitemShut {NoStop}%
\bibitem [{\citenamefont {{Edelman}}\ \emph {et~al.}(2023)\citenamefont
  {{Edelman}}, \citenamefont {{Farr}},\ and\ \citenamefont
  {{Doctor}}}]{2023ApJ...946...16E}%
  \BibitemOpen
  \bibfield  {author} {\bibinfo {author} {\bibfnamefont {B.}~\bibnamefont
  {{Edelman}}}, \bibinfo {author} {\bibfnamefont {B.}~\bibnamefont {{Farr}}}, \
  and\ \bibinfo {author} {\bibfnamefont {Z.}~\bibnamefont {{Doctor}}},\ }\href
  {\doibase 10.3847/1538-4357/acb5ed} {\bibfield  {journal} {\bibinfo
  {journal} {\apj}\ }\textbf {\bibinfo {volume} {946}},\ \bibinfo {eid} {16}
  (\bibinfo {year} {2023})},\ \Eprint {http://arxiv.org/abs/2210.12834}
  {arXiv:2210.12834 [astro-ph.HE]} \BibitemShut {NoStop}%
\bibitem [{\citenamefont {{Tong}}\ \emph {et~al.}(2026)\citenamefont {{Tong}},
  \citenamefont {{Fishbach}}, \citenamefont {{Thrane}}, \citenamefont
  {{Mould}}, \citenamefont {{Callister}}, \citenamefont {{Farah}},
  \citenamefont {{Guttman}}, \citenamefont {{Banagiri}}, \citenamefont
  {{Beltran-Martinez}}, \citenamefont {{Farr}}, \citenamefont {{Galaudage}},
  \citenamefont {{Godfrey}}, \citenamefont {{Heinzel}}, \citenamefont
  {{Kalomenopoulos}}, \citenamefont {{Miller}},\ and\ \citenamefont
  {{Vijaykumar}}}]{2026Natur.652..874T}%
  \BibitemOpen
  \bibfield  {author} {\bibinfo {author} {\bibfnamefont {H.}~\bibnamefont
  {{Tong}}}, \bibinfo {author} {\bibfnamefont {M.}~\bibnamefont {{Fishbach}}},
  \bibinfo {author} {\bibfnamefont {E.}~\bibnamefont {{Thrane}}}, \bibinfo
  {author} {\bibfnamefont {M.}~\bibnamefont {{Mould}}}, \bibinfo {author}
  {\bibfnamefont {T.~A.}\ \bibnamefont {{Callister}}}, \bibinfo {author}
  {\bibfnamefont {A.~M.}\ \bibnamefont {{Farah}}}, \bibinfo {author}
  {\bibfnamefont {N.}~\bibnamefont {{Guttman}}}, \bibinfo {author}
  {\bibfnamefont {S.}~\bibnamefont {{Banagiri}}}, \bibinfo {author}
  {\bibfnamefont {D.}~\bibnamefont {{Beltran-Martinez}}}, \bibinfo {author}
  {\bibfnamefont {B.}~\bibnamefont {{Farr}}}, \bibinfo {author} {\bibfnamefont
  {S.}~\bibnamefont {{Galaudage}}}, \bibinfo {author} {\bibfnamefont
  {J.}~\bibnamefont {{Godfrey}}}, \bibinfo {author} {\bibfnamefont
  {J.}~\bibnamefont {{Heinzel}}}, \bibinfo {author} {\bibfnamefont
  {M.}~\bibnamefont {{Kalomenopoulos}}}, \bibinfo {author} {\bibfnamefont
  {S.~J.}\ \bibnamefont {{Miller}}}, \ and\ \bibinfo {author} {\bibfnamefont
  {A.}~\bibnamefont {{Vijaykumar}}},\ }\href {\doibase
  10.1038/s41586-026-10359-0} {\bibfield  {journal} {\bibinfo  {journal}
  {\nat}\ }\textbf {\bibinfo {volume} {652}},\ \bibinfo {pages} {874} (\bibinfo
  {year} {2026})},\ \Eprint {http://arxiv.org/abs/2509.04151} {arXiv:2509.04151
  [astro-ph.HE]} \BibitemShut {NoStop}%
\bibitem [{\citenamefont {{Antonini}}\ \emph
  {et~al.}(2025{\natexlab{b}})\citenamefont {{Antonini}}, \citenamefont
  {{Romero-Shaw}}, \citenamefont {{Callister}}, \citenamefont {{Dosopoulou}},
  \citenamefont {{Chattopadhyay}}, \citenamefont {{Gieles}},\ and\
  \citenamefont {{Mapelli}}}]{2025arXiv250904637A}%
  \BibitemOpen
  \bibfield  {author} {\bibinfo {author} {\bibfnamefont {F.}~\bibnamefont
  {{Antonini}}}, \bibinfo {author} {\bibfnamefont {I.}~\bibnamefont
  {{Romero-Shaw}}}, \bibinfo {author} {\bibfnamefont {T.}~\bibnamefont
  {{Callister}}}, \bibinfo {author} {\bibfnamefont {F.}~\bibnamefont
  {{Dosopoulou}}}, \bibinfo {author} {\bibfnamefont {D.}~\bibnamefont
  {{Chattopadhyay}}}, \bibinfo {author} {\bibfnamefont {M.}~\bibnamefont
  {{Gieles}}}, \ and\ \bibinfo {author} {\bibfnamefont {M.}~\bibnamefont
  {{Mapelli}}},\ }\href {\doibase 10.48550/arXiv.2509.04637} {\bibfield
  {journal} {\bibinfo  {journal} {arXiv e-prints}\ ,\ \bibinfo {eid}
  {arXiv:2509.04637}} (\bibinfo {year} {2025}{\natexlab{b}})},\ \Eprint
  {http://arxiv.org/abs/2509.04637} {arXiv:2509.04637 [astro-ph.HE]}
  \BibitemShut {NoStop}%
\bibitem [{\citenamefont {{Afroz}}\ and\ \citenamefont
  {{Mukherjee}}(2025)}]{2025arXiv250909123A}%
  \BibitemOpen
  \bibfield  {author} {\bibinfo {author} {\bibfnamefont {S.}~\bibnamefont
  {{Afroz}}}\ and\ \bibinfo {author} {\bibfnamefont {S.}~\bibnamefont
  {{Mukherjee}}},\ }\href {\doibase 10.48550/arXiv.2509.09123} {\bibfield
  {journal} {\bibinfo  {journal} {arXiv e-prints}\ ,\ \bibinfo {eid}
  {arXiv:2509.09123}} (\bibinfo {year} {2025})},\ \Eprint
  {http://arxiv.org/abs/2509.09123} {arXiv:2509.09123 [astro-ph.HE]}
  \BibitemShut {NoStop}%
\bibitem [{\citenamefont {{Adamcewicz}}\ \emph {et~al.}(2025)\citenamefont
  {{Adamcewicz}}, \citenamefont {{Guttman}}, \citenamefont {{Lasky}},\ and\
  \citenamefont {{Thrane}}}]{2025ApJ...994..261A}%
  \BibitemOpen
  \bibfield  {author} {\bibinfo {author} {\bibfnamefont {C.}~\bibnamefont
  {{Adamcewicz}}}, \bibinfo {author} {\bibfnamefont {N.}~\bibnamefont
  {{Guttman}}}, \bibinfo {author} {\bibfnamefont {P.~D.}\ \bibnamefont
  {{Lasky}}}, \ and\ \bibinfo {author} {\bibfnamefont {E.}~\bibnamefont
  {{Thrane}}},\ }\href {\doibase 10.3847/1538-4357/ae1370} {\bibfield
  {journal} {\bibinfo  {journal} {\apj}\ }\textbf {\bibinfo {volume} {994}},\
  \bibinfo {eid} {261} (\bibinfo {year} {2025})},\ \Eprint
  {http://arxiv.org/abs/2509.04706} {arXiv:2509.04706 [astro-ph.HE]}
  \BibitemShut {NoStop}%
\bibitem [{\citenamefont {{Banagiri}}\ \emph {et~al.}(2025)\citenamefont
  {{Banagiri}}, \citenamefont {{Thrane}},\ and\ \citenamefont
  {{Lasky}}}]{2025arXiv250915646B}%
  \BibitemOpen
  \bibfield  {author} {\bibinfo {author} {\bibfnamefont {S.}~\bibnamefont
  {{Banagiri}}}, \bibinfo {author} {\bibfnamefont {E.}~\bibnamefont
  {{Thrane}}}, \ and\ \bibinfo {author} {\bibfnamefont {P.~D.}\ \bibnamefont
  {{Lasky}}},\ }\href {\doibase 10.48550/arXiv.2509.15646} {\bibfield
  {journal} {\bibinfo  {journal} {arXiv e-prints}\ ,\ \bibinfo {eid}
  {arXiv:2509.15646}} (\bibinfo {year} {2025})},\ \Eprint
  {http://arxiv.org/abs/2509.15646} {arXiv:2509.15646 [astro-ph.HE]}
  \BibitemShut {NoStop}%
\bibitem [{\citenamefont {{Gerosa}}\ and\ \citenamefont
  {{Berti}}(2017)}]{2017PhRvD..95l4046G}%
  \BibitemOpen
  \bibfield  {author} {\bibinfo {author} {\bibfnamefont {D.}~\bibnamefont
  {{Gerosa}}}\ and\ \bibinfo {author} {\bibfnamefont {E.}~\bibnamefont
  {{Berti}}},\ }\href {\doibase 10.1103/PhysRevD.95.124046} {\bibfield
  {journal} {\bibinfo  {journal} {\prd}\ }\textbf {\bibinfo {volume} {95}},\
  \bibinfo {eid} {124046} (\bibinfo {year} {2017})},\ \Eprint
  {http://arxiv.org/abs/1703.06223} {arXiv:1703.06223 [gr-qc]} \BibitemShut
  {NoStop}%
\bibitem [{\citenamefont {Wang}\ \emph {et~al.}(2026)\citenamefont {Wang},
  \citenamefont {Li}, \citenamefont {Gao}, \citenamefont {Tang},\ and\
  \citenamefont {Fan}}]{Wang:2025nhf}%
  \BibitemOpen
  \bibfield  {author} {\bibinfo {author} {\bibfnamefont {Y.-Z.}\ \bibnamefont
  {Wang}}, \bibinfo {author} {\bibfnamefont {Y.-J.}\ \bibnamefont {Li}},
  \bibinfo {author} {\bibfnamefont {S.-J.}\ \bibnamefont {Gao}}, \bibinfo
  {author} {\bibfnamefont {S.-P.}\ \bibnamefont {Tang}}, \ and\ \bibinfo
  {author} {\bibfnamefont {Y.-Z.}\ \bibnamefont {Fan}},\ }\href {\doibase
  10.1007/s11433-026-3004-6} {\bibfield  {journal} {\bibinfo  {journal} {Sci.
  China Phys. Mech. Astron.}\ }\textbf {\bibinfo {volume} {69}},\ \bibinfo
  {pages} {299562} (\bibinfo {year} {2026})},\ \Eprint
  {http://arxiv.org/abs/2510.22698} {arXiv:2510.22698 [astro-ph.HE]}
  \BibitemShut {NoStop}%
\bibitem [{\citenamefont {{Woosley}}\ and\ \citenamefont
  {{Heger}}(2021)}]{2021ApJ...912L..31W}%
  \BibitemOpen
  \bibfield  {author} {\bibinfo {author} {\bibfnamefont {S.~E.}\ \bibnamefont
  {{Woosley}}}\ and\ \bibinfo {author} {\bibfnamefont {A.}~\bibnamefont
  {{Heger}}},\ }\href {\doibase 10.3847/2041-8213/abf2c4} {\bibfield  {journal}
  {\bibinfo  {journal} {\apjl}\ }\textbf {\bibinfo {volume} {912}},\ \bibinfo
  {eid} {L31} (\bibinfo {year} {2021})},\ \Eprint
  {http://arxiv.org/abs/2103.07933} {arXiv:2103.07933 [astro-ph.SR]}
  \BibitemShut {NoStop}%
\bibitem [{\citenamefont {{Rodriguez}}\ \emph {et~al.}(2016)\citenamefont
  {{Rodriguez}}, \citenamefont {{Zevin}}, \citenamefont {{Pankow}},
  \citenamefont {{Kalogera}},\ and\ \citenamefont
  {{Rasio}}}]{2016ApJ...832L...2R}%
  \BibitemOpen
  \bibfield  {author} {\bibinfo {author} {\bibfnamefont {C.~L.}\ \bibnamefont
  {{Rodriguez}}}, \bibinfo {author} {\bibfnamefont {M.}~\bibnamefont
  {{Zevin}}}, \bibinfo {author} {\bibfnamefont {C.}~\bibnamefont {{Pankow}}},
  \bibinfo {author} {\bibfnamefont {V.}~\bibnamefont {{Kalogera}}}, \ and\
  \bibinfo {author} {\bibfnamefont {F.~A.}\ \bibnamefont {{Rasio}}},\ }\href
  {\doibase 10.3847/2041-8205/832/1/L2} {\bibfield  {journal} {\bibinfo
  {journal} {\apjl}\ }\textbf {\bibinfo {volume} {832}},\ \bibinfo {eid} {L2}
  (\bibinfo {year} {2016})},\ \Eprint {http://arxiv.org/abs/1609.05916}
  {arXiv:1609.05916 [astro-ph.HE]} \BibitemShut {NoStop}%
\bibitem [{\citenamefont {{Li}}\ \emph
  {et~al.}(2024{\natexlab{b}})\citenamefont {{Li}}, \citenamefont {{Tang}},
  \citenamefont {{Gao}}, \citenamefont {{Wu}},\ and\ \citenamefont
  {{Wang}}}]{2024ApJ...977...67L}%
  \BibitemOpen
  \bibfield  {author} {\bibinfo {author} {\bibfnamefont {Y.-J.}\ \bibnamefont
  {{Li}}}, \bibinfo {author} {\bibfnamefont {S.-P.}\ \bibnamefont {{Tang}}},
  \bibinfo {author} {\bibfnamefont {S.-J.}\ \bibnamefont {{Gao}}}, \bibinfo
  {author} {\bibfnamefont {D.-C.}\ \bibnamefont {{Wu}}}, \ and\ \bibinfo
  {author} {\bibfnamefont {Y.-Z.}\ \bibnamefont {{Wang}}},\ }\href {\doibase
  10.3847/1538-4357/ad83b5} {\bibfield  {journal} {\bibinfo  {journal} {\apj}\
  }\textbf {\bibinfo {volume} {977}},\ \bibinfo {eid} {67} (\bibinfo {year}
  {2024}{\natexlab{b}})},\ \Eprint {http://arxiv.org/abs/2404.09668}
  {arXiv:2404.09668 [astro-ph.HE]} \BibitemShut {NoStop}%
\bibitem [{\citenamefont {{Yang}}\ \emph
  {et~al.}(2019{\natexlab{b}})\citenamefont {{Yang}}, \citenamefont {{Bartos}},
  \citenamefont {{Haiman}}, \citenamefont {{Kocsis}}, \citenamefont
  {{M{\'a}rka}}, \citenamefont {{Stone}},\ and\ \citenamefont
  {{M{\'a}rka}}}]{2019ApJ...876..122Y}%
  \BibitemOpen
  \bibfield  {author} {\bibinfo {author} {\bibfnamefont {Y.}~\bibnamefont
  {{Yang}}}, \bibinfo {author} {\bibfnamefont {I.}~\bibnamefont {{Bartos}}},
  \bibinfo {author} {\bibfnamefont {Z.}~\bibnamefont {{Haiman}}}, \bibinfo
  {author} {\bibfnamefont {B.}~\bibnamefont {{Kocsis}}}, \bibinfo {author}
  {\bibfnamefont {Z.}~\bibnamefont {{M{\'a}rka}}}, \bibinfo {author}
  {\bibfnamefont {N.~C.}\ \bibnamefont {{Stone}}}, \ and\ \bibinfo {author}
  {\bibfnamefont {S.}~\bibnamefont {{M{\'a}rka}}},\ }\href {\doibase
  10.3847/1538-4357/ab16e3} {\bibfield  {journal} {\bibinfo  {journal} {\apj}\
  }\textbf {\bibinfo {volume} {876}},\ \bibinfo {eid} {122} (\bibinfo {year}
  {2019}{\natexlab{b}})},\ \Eprint {http://arxiv.org/abs/1903.01405}
  {arXiv:1903.01405 [astro-ph.HE]} \BibitemShut {NoStop}%
\bibitem [{\citenamefont {{Gayathri}}\ \emph {et~al.}(2021)\citenamefont
  {{Gayathri}}, \citenamefont {{Yang}}, \citenamefont {{Tagawa}}, \citenamefont
  {{Haiman}},\ and\ \citenamefont {{Bartos}}}]{2021ApJ...920L..42G}%
  \BibitemOpen
  \bibfield  {author} {\bibinfo {author} {\bibfnamefont {V.}~\bibnamefont
  {{Gayathri}}}, \bibinfo {author} {\bibfnamefont {Y.}~\bibnamefont {{Yang}}},
  \bibinfo {author} {\bibfnamefont {H.}~\bibnamefont {{Tagawa}}}, \bibinfo
  {author} {\bibfnamefont {Z.}~\bibnamefont {{Haiman}}}, \ and\ \bibinfo
  {author} {\bibfnamefont {I.}~\bibnamefont {{Bartos}}},\ }\href {\doibase
  10.3847/2041-8213/ac2cc1} {\bibfield  {journal} {\bibinfo  {journal} {\apjl}\
  }\textbf {\bibinfo {volume} {920}},\ \bibinfo {eid} {L42} (\bibinfo {year}
  {2021})},\ \Eprint {http://arxiv.org/abs/2104.10253} {arXiv:2104.10253
  [gr-qc]} \BibitemShut {NoStop}%
\bibitem [{\citenamefont {{Zhu}}\ and\ \citenamefont
  {{Chen}}(2025)}]{2025ApJ...989L..15Z}%
  \BibitemOpen
  \bibfield  {author} {\bibinfo {author} {\bibfnamefont {L.-G.}\ \bibnamefont
  {{Zhu}}}\ and\ \bibinfo {author} {\bibfnamefont {X.}~\bibnamefont {{Chen}}},\
  }\href {\doibase 10.3847/2041-8213/adf31f} {\bibfield  {journal} {\bibinfo
  {journal} {\apjl}\ }\textbf {\bibinfo {volume} {989}},\ \bibinfo {eid} {L15}
  (\bibinfo {year} {2025})},\ \Eprint {http://arxiv.org/abs/2505.02924}
  {arXiv:2505.02924 [astro-ph.HE]} \BibitemShut {NoStop}%
\bibitem [{\citenamefont {{Wang}}\ \emph {et~al.}(2025)\citenamefont {{Wang}},
  \citenamefont {{Hu}}, \citenamefont {{Chen}}, \citenamefont {{Songsheng}},
  \citenamefont {{Wang}}, \citenamefont {{Zhang}}, \citenamefont {{Du}},
  \citenamefont {{Li}}, \citenamefont {{Luo}}, \citenamefont {{Brotherton}},
  \citenamefont {{Bai}}, \citenamefont {{Guo}}, \citenamefont {{Yang}},
  \citenamefont {{Yao}},\ and\ \citenamefont
  {{Aceituno}}}]{2025arXiv251107716W}%
  \BibitemOpen
  \bibfield  {author} {\bibinfo {author} {\bibfnamefont {J.-M.}\ \bibnamefont
  {{Wang}}}, \bibinfo {author} {\bibfnamefont {C.}~\bibnamefont {{Hu}}},
  \bibinfo {author} {\bibfnamefont {Y.-J.}\ \bibnamefont {{Chen}}}, \bibinfo
  {author} {\bibfnamefont {Y.-Y.}\ \bibnamefont {{Songsheng}}}, \bibinfo
  {author} {\bibfnamefont {Y.-L.}\ \bibnamefont {{Wang}}}, \bibinfo {author}
  {\bibfnamefont {H.}~\bibnamefont {{Zhang}}}, \bibinfo {author} {\bibfnamefont
  {P.}~\bibnamefont {{Du}}}, \bibinfo {author} {\bibfnamefont {Y.-R.}\
  \bibnamefont {{Li}}}, \bibinfo {author} {\bibfnamefont {B.}~\bibnamefont
  {{Luo}}}, \bibinfo {author} {\bibfnamefont {M.~S.}\ \bibnamefont
  {{Brotherton}}}, \bibinfo {author} {\bibfnamefont {J.-M.}\ \bibnamefont
  {{Bai}}}, \bibinfo {author} {\bibfnamefont {W.-J.}\ \bibnamefont {{Guo}}},
  \bibinfo {author} {\bibfnamefont {S.}~\bibnamefont {{Yang}}}, \bibinfo
  {author} {\bibfnamefont {Z.-H.}\ \bibnamefont {{Yao}}}, \ and\ \bibinfo
  {author} {\bibfnamefont {J.}~\bibnamefont {{Aceituno}}},\ }\href {\doibase
  10.48550/arXiv.2511.07716} {\bibfield  {journal} {\bibinfo  {journal} {arXiv
  e-prints}\ ,\ \bibinfo {eid} {arXiv:2511.07716}} (\bibinfo {year} {2025})},\
  \Eprint {http://arxiv.org/abs/2511.07716} {arXiv:2511.07716 [astro-ph.GA]}
  \BibitemShut {NoStop}%
\bibitem [{\citenamefont {{Bartos}}\ and\ \citenamefont
  {{Haiman}}(2026)}]{2026ApJ...996L..44B}%
  \BibitemOpen
  \bibfield  {author} {\bibinfo {author} {\bibfnamefont {I.}~\bibnamefont
  {{Bartos}}}\ and\ \bibinfo {author} {\bibfnamefont {Z.}~\bibnamefont
  {{Haiman}}},\ }\href {\doibase 10.3847/2041-8213/ae2bff} {\bibfield
  {journal} {\bibinfo  {journal} {\apjl}\ }\textbf {\bibinfo {volume} {996}},\
  \bibinfo {eid} {L44} (\bibinfo {year} {2026})},\ \Eprint
  {http://arxiv.org/abs/2508.08558} {arXiv:2508.08558 [astro-ph.HE]}
  \BibitemShut {NoStop}%
\bibitem [{\citenamefont {{Godfrey}}\ \emph {et~al.}(2023)\citenamefont
  {{Godfrey}}, \citenamefont {{Edelman}},\ and\ \citenamefont
  {{Farr}}}]{2023arXiv230401288G}%
  \BibitemOpen
  \bibfield  {author} {\bibinfo {author} {\bibfnamefont {J.}~\bibnamefont
  {{Godfrey}}}, \bibinfo {author} {\bibfnamefont {B.}~\bibnamefont
  {{Edelman}}}, \ and\ \bibinfo {author} {\bibfnamefont {B.}~\bibnamefont
  {{Farr}}},\ }\href {\doibase 10.48550/arXiv.2304.01288} {\bibfield  {journal}
  {\bibinfo  {journal} {arXiv e-prints}\ ,\ \bibinfo {eid} {arXiv:2304.01288}}
  (\bibinfo {year} {2023})},\ \Eprint {http://arxiv.org/abs/2304.01288}
  {arXiv:2304.01288 [astro-ph.HE]} \BibitemShut {NoStop}%
\bibitem [{\citenamefont {{Xue}}\ \emph {et~al.}(2025)\citenamefont {{Xue}},
  \citenamefont {{Tagawa}}, \citenamefont {{Haiman}},\ and\ \citenamefont
  {{Bartos}}}]{2025PhRvD.112f3034X}%
  \BibitemOpen
  \bibfield  {author} {\bibinfo {author} {\bibfnamefont {L.}~\bibnamefont
  {{Xue}}}, \bibinfo {author} {\bibfnamefont {H.}~\bibnamefont {{Tagawa}}},
  \bibinfo {author} {\bibfnamefont {Z.}~\bibnamefont {{Haiman}}}, \ and\
  \bibinfo {author} {\bibfnamefont {I.}~\bibnamefont {{Bartos}}},\ }\href
  {\doibase 10.1103/5m1n-qh9v} {\bibfield  {journal} {\bibinfo  {journal}
  {\prd}\ }\textbf {\bibinfo {volume} {112}},\ \bibinfo {eid} {063034}
  (\bibinfo {year} {2025})},\ \Eprint {http://arxiv.org/abs/2504.19570}
  {arXiv:2504.19570 [astro-ph.HE]} \BibitemShut {NoStop}%
\end{thebibliography}%

\clearpage

\appendix

\renewcommand{\thesection}{\Roman{section}} 
\renewcommand{\appendixname}{} 

\setcounter{figure}{0}
\renewcommand\thefigure{S\arabic{figure}}
\setcounter{table}{0}
\renewcommand\thetable{S\arabic{table}}

\section*{Supplemental Material}

\section{Hierarchical Bayesian Inference} \label{sec:methods}
Given a population distribution $\Lambda$, the likelihood of the GW data $\{d\}$ from $N_{\rm det}$ detections is  \citep{2019MNRAS.486.1086M,
2023PhRvX..13a1048A,
2025arXiv250818083T}, 
\begin{equation}
\mathcal{L}(\{d\}|\Lambda)\propto N^{N_{\rm det}}e^{-N_{\rm exp}} \prod_{i}^{N_{\rm det}}{\int{\pi(\theta_i|\Lambda)\mathcal{L}(d_i|\theta_i)d\theta_i}},
\end{equation}
where $N=\int{R(z|\Lambda)\frac{dV_{\rm c}}{dz}\frac{T_{\rm obs}}{1+z}dz}$ is the total number of mergers in the surveyed time-space volume, and $N_{\rm exp}=N\int{P({\rm det}|\theta)\pi(\theta|\Lambda)d\theta}$ is the expected number of detections, with detection probability $P({\rm det}|\theta)$. $N_{\rm exp}$ is calculated using a Monte Carlo integral over the referred injection \citep{2025PhRvD.112j2001E} (Adopted from https://zenodo.org/records/16740128), and $\mathcal{L}(d_i|\theta_i)$ is evaluated using the posterior samples (Posterior samples \citep{
2024PhRvD.109b2001A,2023PhRvX..13d1039A,2025arXiv250818082T} are taken from \href{https://gwosc.org/eventapi/html/GWTC/}{GWOSC}).
Following \citep{2025arXiv250818083T}, we constrain the total uncertainties of the Monte Carlo integrals $\sigma_{\rm tot} < 1$ to ensure accurate evaluation of likelihood, and choose a threshold of false-alarm rate (FAR) of $<1~yr^{-1}$ to select the events, and exclude the GW190814 for its unclear nature \citep{
2023PhRvX..13a1048A}, so that 153 events are selected for our analysis. The  posterior samples used in our analysis are the same as that of \citet{2025arXiv250818083T}. We use the Pymultinest sampler \citep{2016ascl.soft06005B}, to obtain the posterior distribution of the hyperparameters.

\section{Population model}\label{sec:models}
As already mentioned in the main text, our base $m-\chi-\cos\theta$ population model is
$\pi(m,\chi,\cos\theta|\Lambda)=\sum_{i=1,2} P_{\mathcal{PS}}(m|\Lambda_i) \times P_{\mathcal{S}}(\chi |\Lambda_i) \times P_{\mathcal{S}}(\cos\theta |\Lambda_i) \times r_i$, 
where $P_{\mathcal{PS}}$ and $P_{\mathcal{S}}$ are the $PowerLaw+Spline$ and $Spline$ models for the distributions of component masses and spin properties. 
These functions are flexible (i.e. weakly modeled) so that the obtained distributions will be data-driven \citep{2022ApJ...924..101E,2023ApJ...946...16E}, which are expressed as
\begin{equation}
P_{\mathcal{PS},i}(m)\propto m^{-\alpha_i} e^{f(m | \{f_j^i\}_{j=1}^{12})} {\rm SF}(m | m_{{\rm min},i}, \delta_i) [m_{{\rm min},i}, m_{{\rm max},i}],
\end{equation}

\begin{equation}
P_{\mathcal{S},i}(\chi)\propto e^{n(\chi| \{n_j^i\}_{j=1}^{5})}[\chi_{{\rm min},i}, \chi_{{\rm max},i}],
\end{equation}
and 
\begin{equation}
P_{\mathcal{S},i}(\cos\theta)\propto e^{t(\cos\theta|\{t_j^i\}_{j=1}^{4})}[-1,1].
\end{equation} 
where ${\rm SF}(m | m_{{\rm min},i}, \delta_i)$ is the smooth function as defined in  \citet{2021ApJ...913L...7A}.
For the mass distribution, 12 knots to interpolate the perturbation function are linearly set in log space between $[6,80]M_{\odot}$. While 5 (4) knots are set linearly in $[0,1]$ ($[-1,1]$) for the perturbation function $n(\chi)$ ($t(\cos\theta)$) of $\chi$ ($\cos\theta$) distribution.

The total population model takes the form of 
$\pi(\lambda|\Lambda)\propto \pi(m_1,\chi_1,\cos\theta_1|\Lambda)\times \pi(m_2,\chi_2,\cos\theta_2|\Lambda) \times P_{\rm pair}(q|\Lambda) \times \pi_{\rm z}(z|\gamma)$, 
with 
\begin{equation}
P_{\rm pair}(q|\Lambda) \propto q^{\beta u(q|\{u^q_j\}_{j=1}^{4})} \times \Theta(q<1),
\end{equation}
where $f_\mathcal{S}(q)$ is the perturbation of paring function beyond a simple power law, which is interpolated with 4 knots linearly in [0,1]. The redshift distribution is
\begin{equation}\label{eq:z}
\pi_{\rm z}(z|\gamma) \propto \frac{{\rm d}V_c}{(1+z){\rm d}z} \times (1+z)^\gamma.
\end{equation}
All the hyperparameters are summarized in Table~\ref{tab:prior}.

\begin{table*}[htpb]
\centering
\caption{Summary of model parameters.}\label{tab:prior}
\begin{tabular}{lcccc} 
\hline
\hline
Parameter     &  Description & Prior \\
\hline
$m_{\rm min,i}[M_{\odot}]$   & The minimum mass & $U(2,50)$  \\
$m_{\rm max,1}/m_{\rm max,2}[M_{\odot}]$   & The maximum mass & $U(20,100)$/$U(20,200)$  \\
$\alpha_i$ & Slope index of the power-law mass function & $U(-8,8)$ \\
$\delta_{\rm m,1}$ / $\delta_{\rm m,2}$ $[M_{\odot}]$ & Smooth scale of the mass lower edge & $U(0,10)$ / $U(0,20)$ \\
$\{f_j^i\}_{j=2}^{11}$ & Interpolation values of perturbation function & $\mathcal{N}(0,1)$ \\
$r_i$ & mixture fraction for the $i$-th subpopulation & $Dir$ \\
constraints & & $m_{\rm min,i}<m_{\rm max,i}-20$ \\
$\beta_{q}$ & Slope index of the pairing function & $U(-8,8)$ \\
$\{u_j^q\}_{j=1}^{4}$ & perturbation values for the pairing function& $\mathcal{N}(0,1)$ \\
$\chi_{{\rm min},1}$ / $\chi_{{\rm min},2}$   & Lower edge for $\chi$ distribution  & 0 / $U(0,0.8)$  \\
$\chi_{{\rm max},1}$/ $\chi_{{\rm max},2}$   & Upper edge for $\chi$ distribution &  $U(0.2,1)$ /1   \\
$\{n_j^i\}_{j=1}^{5}$ & perturbation values for $\chi$ distribution& $\mathcal{N}(0,2)$ \\
$\{t_j^i\}_{j=1}^{4}$ & perturbation values for $\cos\theta$ distribution& $\mathcal{N}(0,1)$ \\
$\lg (R_0[{\rm Gpc}^{-3}~{\rm yr}^{-1}])$ & Local merger rate  density & $ U(-3,3)$ \\
$\gamma$ & Slope of the power-law & $U(-2,7)$ \\
\hline
\hline
\end{tabular}
\\
\begin{tabular}{l}
Note: $U$,  $Dir$, and $\mathcal{N}$ are for Uniform, Dirichlet, and Normal distribution.
\end{tabular}
\end{table*} 

With the flexible nonparametric population model (note that in \citet{2024PhRvL.133e1401L} the spin distributions are modeled with parametric formula), we successfully identify two subpopulations of BHs by a Bayes factor of $\ln\mathcal{B}_{1pop}^{2pop}=12$, strengthening the finding with GWTC-3 \citep{2024PhRvL.133e1401L}. 
Figure~1. (in the main text) shows the reconstructed component-mass and spin-magnitude distributions of two subpopulations inferred with the nonparametric model introduced above. Figure~1. in the main text displays the population-informed posterior distribution of each event in the $\chi-m$ panel. The two subpopulations are clearly separated, although the maximum mass of the first subpopulation has extend to $m_{\rm max,1}\sim 70~M_{\odot}$ with sizable uncertainty. Such a value is still consistent with the core-collapse BHs of current knowledge of stellar evolution theories, in particular in the presence of rapid rotation \citep{2021ApJ...912L..31W}.
We notice that in some simplified parametric approaches, 
the mass cutoff (or transition points between two subpopulations) can be constrained to tighter values of $\sim 45\pm 5 M_\odot$ with GWTC-4 \citep{2026Natur.652..874T,2025arXiv250904637A,2025arXiv250915646B}, even with GWTC-3 \citep{2022ApJ...941L..39W}. But the emergence of a new group of low-spin but massive ($\sim 50-70M_\odot$) suggests a significantly higher cutoff mass (see also
\citep{Wang:2025nhf}) and hence challenges a low pair-instability mass cutoff.

The component mass function of the BHs shows three peaks at $\sim 10~M_\odot$, $\sim 18~M_\odot$, $\sim 32~M_\odot$, which are consistent with the results in previous works \citep{2021ApJ...917...33L,2024MNRAS.527..298T,2025arXiv250818083T}.
Additionally, we find that the minimum mass of the second subpopulation may extend to $\lesssim 20~M_\odot$. In particular, the primary object of GW231118\_005626 has a large spin and a mass of $\sim18~M_\odot$, which is consistent with being a remnant of a BBH merger with the component masses of $9-10~M_\odot$, suggesting that some of the BBHs at $\sim 10~M_\odot$-peak may be dynamically assembled.

\begin{figure*}
	\centering  
\includegraphics[width=0.98\linewidth]{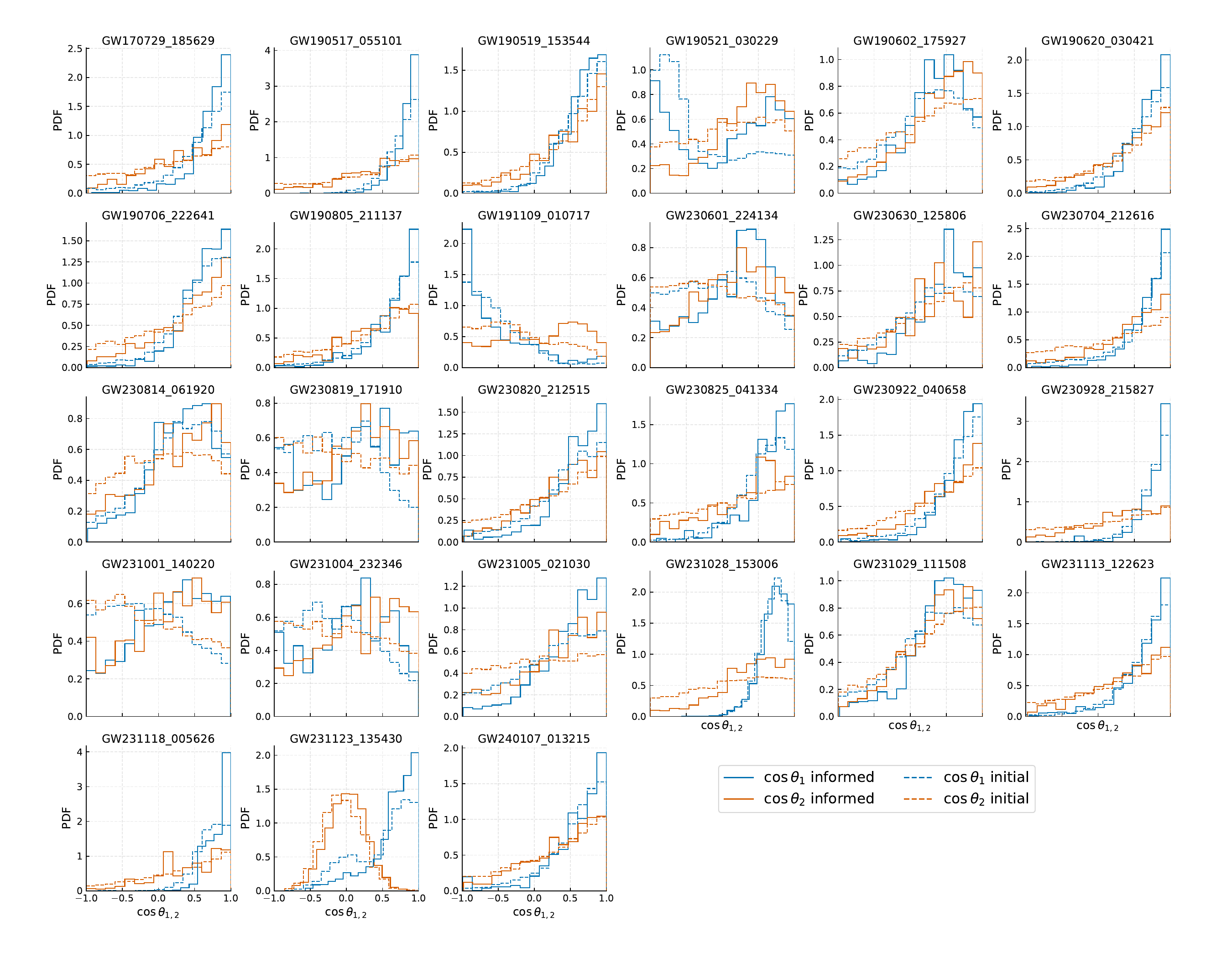}
\caption{{Posterior distributions of $\cos\theta_{1,2}$ for candidate hierarchical merger BBHs (with $\bar{\chi_1}>0.4$ or $\bar{\chi_2}>0.4$) reweighed to a population-informed prior, comparing to that of initial samples obtained with default prior by \citet{2024PhRvD.109b2001A,2023PhRvX..13d1039A} and \citet{2025arXiv250818082T}.}}
\label{app:theta_dist}
\end{figure*}

\begin{figure}
	\centering  
\includegraphics[width=0.98\linewidth]{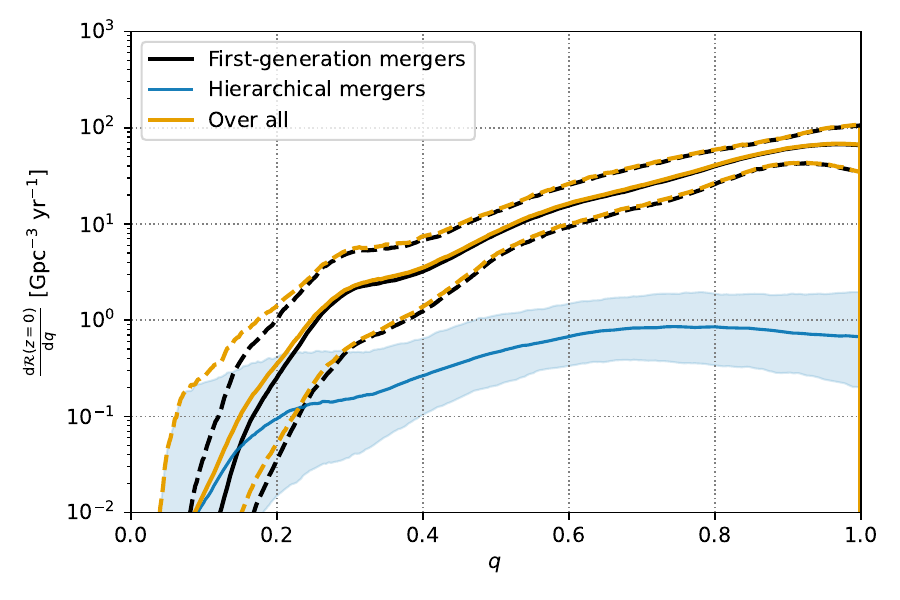}
\caption{{Reconstructed mass-ratio distributions of the first-generation and hierarchical mergers. The solid curves are the medians and the dashed curves are for the 90\% credible intervals.}}
\label{app:q_dist}
\end{figure}

{Figure~\ref{app:theta_dist} presents the $\cos\theta_{1,2}$ distributions of candidate hierarchical mergers (with $\bar{\chi_1}>0.4$ or $\bar{\chi_2}>0.4$). We find that most events (with both primary and secondary BHs) tend to have nearly aligned spins, although $\cos\theta_2$ is not measured as tight as $\cos\theta_1$ for individual event.
Figure~\ref{app:q_dist} shows the mass ratio distributions of the first-generation and hierarchical mergers (containing at least one higher-generation BH in each binary). It shows that the mass-ratio distribution of the hierarchical mergers is flatter than that of the first-generation mergers.
}

\section{Model comparison}\label{sec:compare}

\subsection{Aligned or Isotropic}\label{subsec:aligned}

\begin{table}[htpb]
\centering
\caption{Model comparison}\label{tab:results}
\begin{tabular}{lccc}
\hline
\hline
Model     &  $\ln{\mathcal{B}}$  &  ${\mathcal{B}}$ \\
\hline
${\rm Pop}_2$ Uniform $\cos\theta$  & 0 & 1  \\
${\rm Pop}_2$ Gaussian $\cos\theta$  & 4.5  & 85  \\
Nonparametric model & 5.7 & 298\\
\hline
\end{tabular}
\\
\begin{tabular}{l}
Note: All the (ln) Bayes factors are relative to the model with\\
 isotropic spin for the second subpopulation.
\end{tabular}
\end{table} 

For model comparison, we apply the simple parametric models to describe the spin properties of two subpopulations. 
We concentrate on whether the spin orientations of the second subpopulation are isotropic or aligned. Therefore we set 
\begin{equation}
P_2(\cos\theta) = \mathcal{U}(\cos\theta | -1,1) 
\end{equation}
for the case of isotropic spins, and \begin{equation}
P_2(\cos\theta) = \mathcal{G}(\cos\theta | 1, \sigma_{\rm t,2}, -1,1)
\end{equation}
for the case of aligned spins, where $\mathcal{U}$ is a uniform distribution and $\mathcal{G}$ is a Gaussian distribution, with center value of 1 and width of $\sigma_{\rm t,2}$, truncated in (-1,1).
The spin magnitudes of the two subpopulations are described as $P_1(\chi)=\mathcal{G}( \chi | \mu_{\chi,1} ,\sigma_{\chi,1},0, \chi_{\rm max,1})$ and  $P_2(\chi)=\mathcal{G}( \chi | \mu_{\chi,2} ,\sigma_{\chi,2}, \chi_{\rm min,2}, 1)$. The spin orientation of the first subpopulation is $P_1(\cos\theta)=\mathcal{U}(\cos\theta | -1,1)(1-\zeta_1)+\mathcal{G}(\cos\theta | \mu_{t,1},\sigma_{t,1},-1,1) \zeta_1$. 

\begin{figure}
	\centering  
\includegraphics[width=0.99\linewidth]{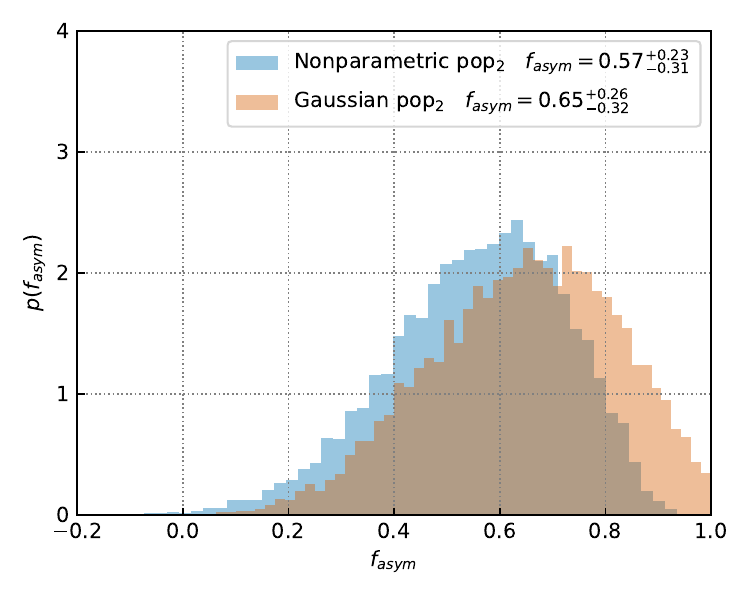}
\caption{Comparison of the $f_{\rm asym}$ of the second subpopulation obtained with the Gaussian model and nonparametric models. It shows the results are consistent between the two models.}
\label{app:asym}
\end{figure}

\begin{figure}
	\centering  
\includegraphics[width=0.98\linewidth]{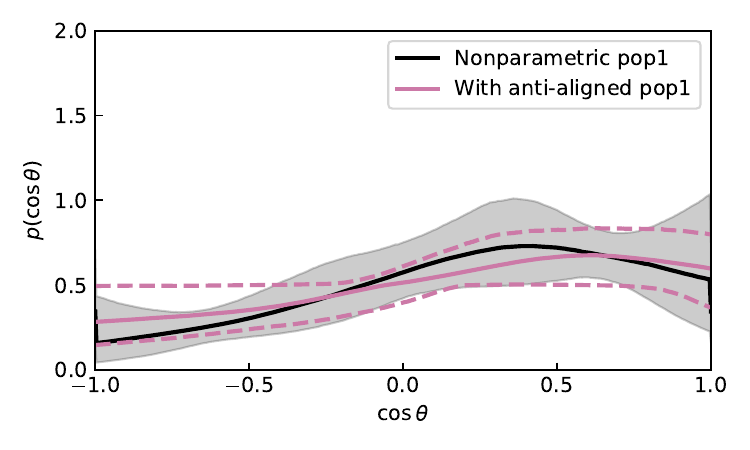}
\includegraphics[width=0.98\linewidth]{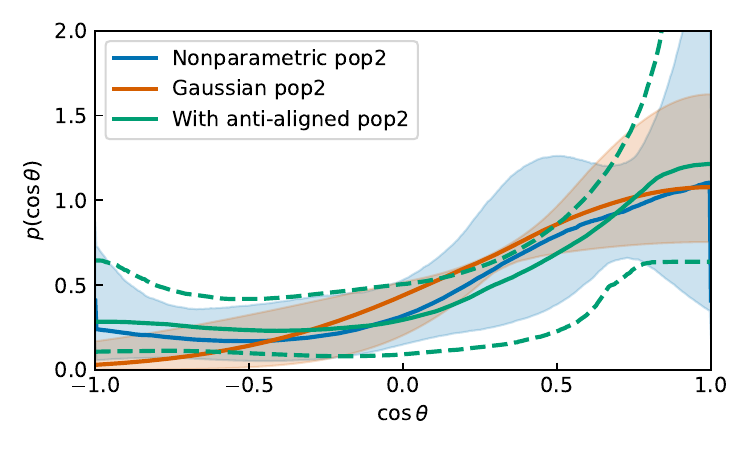}
\caption{The $\cos\theta$ distributions obtained with the parametric models introduced in Section~\ref{sec:compare} compared to that obtained with nonparametric model for the first (top) and second (bottom) subpopulations.}
\label{app:compare_tilt3}
\end{figure}

We find the scenario of isotropic spins (Uniform $\cos\theta$) for the second subpopulation is disfavored compared to the aligned scenario (Gaussian $\cos\theta$), by a Bayes factor of $\ln\mathcal{B}=4.5$. It is even more disfavored compared to the nonparametric model by $\ln\mathcal{B}=5.7$. 
We find that the $\cos\theta$ distribution and the corresponding asymmetric fraction of the second subpopulation, obtained by the Gaussian model, are highly consistent with that obtained by the nonparametric model, see Figure~\ref{app:asym} and  Figure~\ref{app:compare_tilt3}. 

\subsection{Aligned and Anti-aligned assemblies}\label{subsec:anti}
\begin{figure*}
	\centering  
\includegraphics[width=0.8\linewidth]{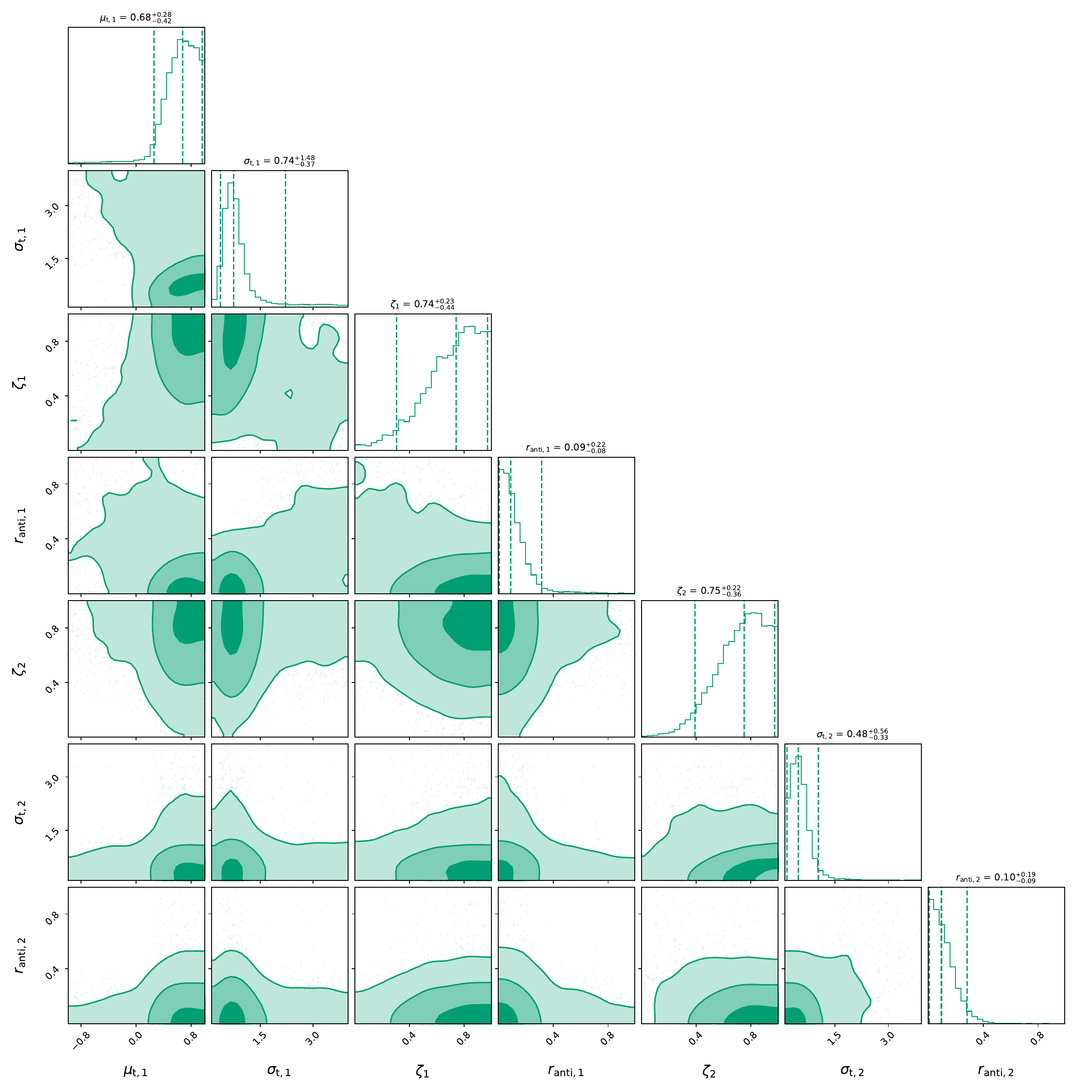}
\caption{The posterior distribution of hyperparameters obtained in model~\ref{model:anti}. The values are for median and 90\% credible intervals. It shows that the anti-aligned fractions are consistent with zero in both subpopulations.}
\label{app:anti_corner}
\end{figure*}

To investigate whether there are isotropic and anti-aligned assemblies in the two subpopulations, we introduce another parametric model for the first and second subpopulations, respectively,
\begin{equation*}\label{model:anti1}
\begin{aligned}
P_1(\cos\theta)=&\mathcal{U}(\cos\theta | -1,1)(1-\zeta_1)+\\
&(\mathcal{G}(\cos\theta | \mu_{\rm t,1} ,\sigma_{t,1},-1,1)(1-r_{\rm anti,1})+\\
&\mathcal{G}(\cos\theta | -1 ,\sigma_{t,2},-1,1)r_{\rm anti,1}) \zeta_1,
\end{aligned}
\end{equation*}
and
\begin{equation*}\label{model:anti}
\begin{aligned}
P_2(\cos\theta)=&\mathcal{U}(\cos\theta | -1,1)(1-\zeta_2)+\\
&(\mathcal{G}(\cos\theta | 1 ,\sigma_{t,2},-1,1)(1-r_{\rm anti,2})+\\
&\mathcal{G}(\cos\theta | -1 ,\sigma_{t,2},-1,1)r_{\rm anti,2}) \zeta_2,
\end{aligned}
\end{equation*}
where $\zeta_1$ ($\zeta_2$) is the non-isotropic fraction, and $r_{\rm anti,1}$ ($r_{\rm anti,2}$) is the mixture fraction of the anti-aligned assembly. Note that we do not fix $\mu_{\rm t,1}=1$ for the first subpopulation, inspired by the results obtained with the nonparametric model. 
We find both $r_{\rm anti,1}$ and $r_{\rm anti,2}$ are consistent with zero, as shown in Figure~\ref{app:anti_corner}, so that an anti-aligned assembly is not necessary within GW data.

It turns out that the reconstructed $\cos\theta$ distribution of the second subpopulation with model~\ref{model:anti} is highly consistent with that obtained with the nonparametric model.
In the main text we report the non-isotropic fraction $f_{\rm non-iso}\gtrsim f_{\rm aysm}=0.57^{+0.23}_{-0.32}$ obtained with the nonparametric model, which is consistent with the non-isotropic fraction defined in this parametric formula $\zeta_2=0.75^{+0.22}_{-0.36}$. The $\zeta_2$ is approaching 1 and $\zeta_2=0$ is almost ruled out, which means that the non-isotropic fraction is dominated, and the fully isotropic scenario is strongly disfavored. 

\section{Impact of degeneracy between parameters}\label{sec:correlation}

\begin{figure*}
	\centering  
\includegraphics[width=0.98\linewidth]{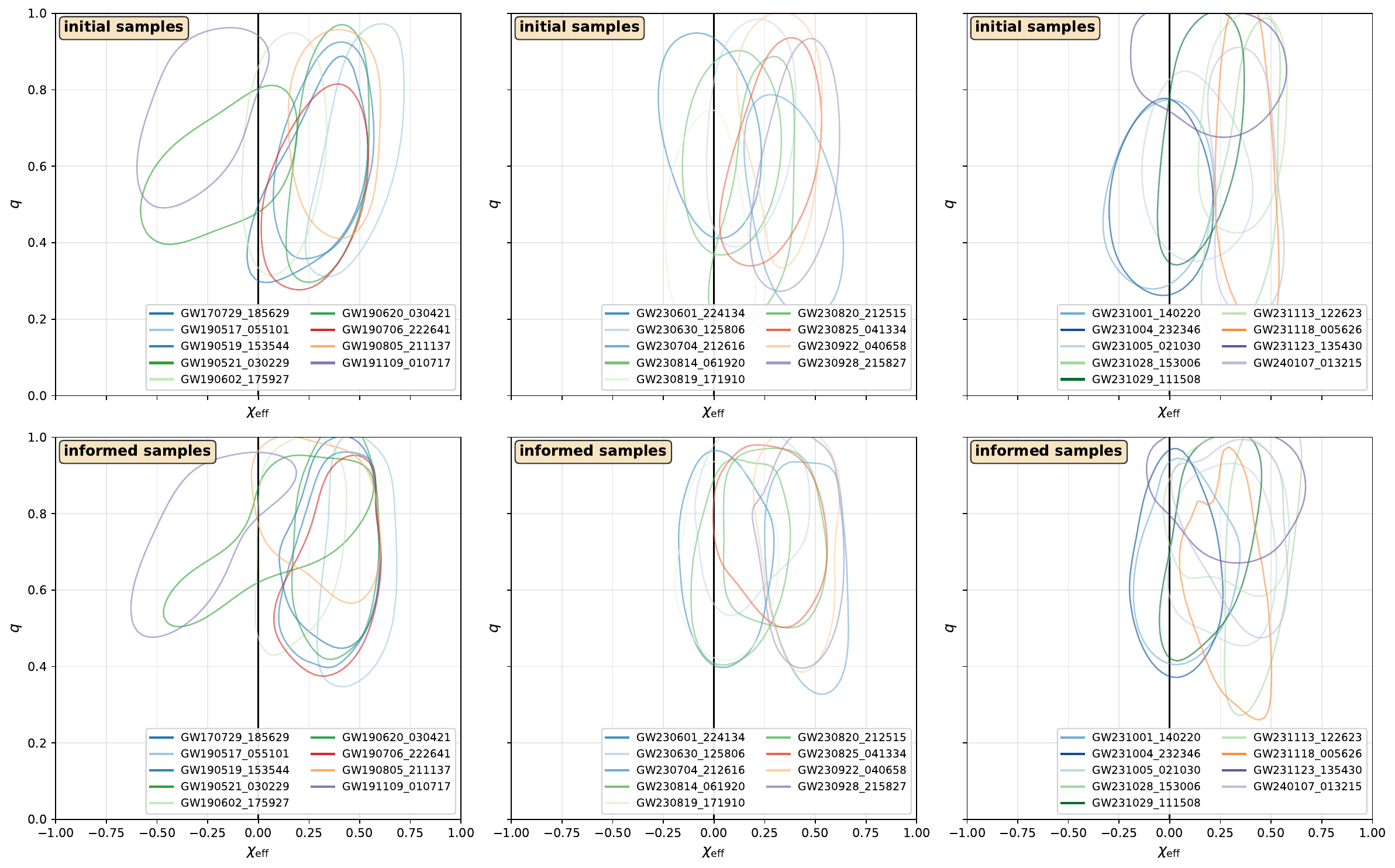}
\caption{{Posterior distributions of $\chi_{\rm eff}-q$ for candidate hierarchical merger BBHs (with $\bar{\chi_1}>0.4$ or $\bar{\chi_2}>0.4$) reweighed to a population-informed prior, comparing to that of initial samples obtained with default prior by \citet{2024PhRvD.109b2001A,2023PhRvX..13d1039A} and \citet{2025arXiv250818082T}.}}
\label{app:chieff_q_dist}
\end{figure*}

\begin{figure*}
	\centering  
\includegraphics[width=0.98\linewidth]{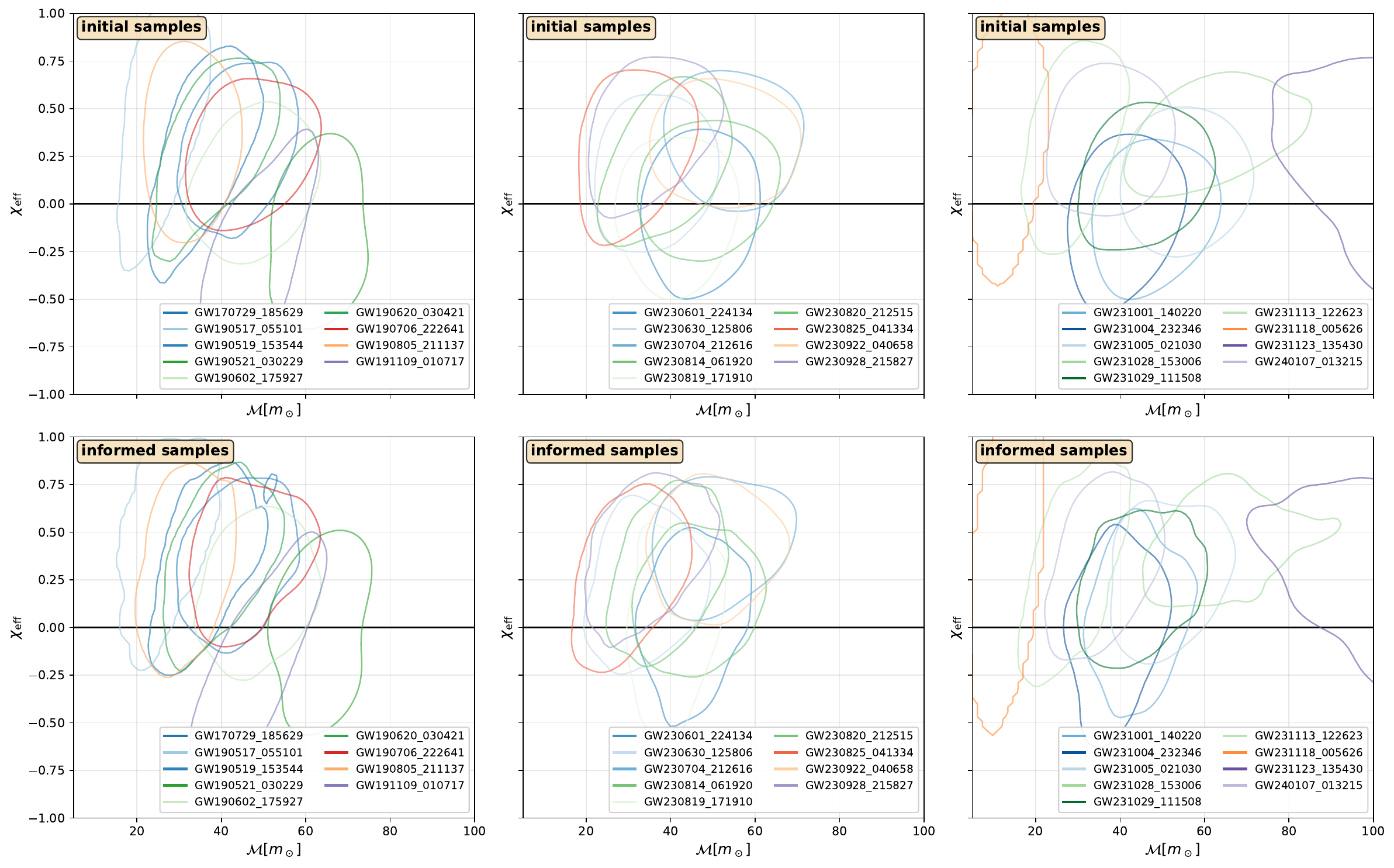}
\caption{{Posterior distributions of chirp mass $\mathcal{M}$ and $\chi_{\rm eff}$ for candidate hierarchical merger BBHs (with $\bar{\chi_1}>0.4$ or $\bar{\chi_2}>0.4$) reweighed to a population-informed prior, comparing to that of initial samples obtained with default prior by \citet{2024PhRvD.109b2001A,2023PhRvX..13d1039A} and \citet{2025arXiv250818082T}.}}
\label{app:Mc_chieff_dist}
\end{figure*}

{Degeneracies exist among the parameters of BBHs, such as correlations between mass ratio and effective spin, and between mass and effective spin \citep{2013PhRvD..87b4035B,2017PhRvD..96l4041W,2019MNRAS.484.4216R}.
Here we examine whether such correlations could affect our results. The effective spin serves as an indicator of BBH alignment and is also known to correlate with mass ratio and chirp mass in parameter estimation \citep{2019MNRAS.484.4216R}. Therefore, we plot the posterior distributions of the candidate hierarchical mergers (indicated as $\bar{\chi_1}>0.4$ or $\bar{\chi_2}>0.4$) in the mass-ratio versus effective-spin and chirp-mass versus effective-spin panels, see Figure~\ref{app:chieff_q_dist} and Figure~\ref{app:Mc_chieff_dist}.
For both the initial samples, obtained with default priors by \citet{2024PhRvD.109b2001A,2023PhRvX..13d1039A} and \citet{2025arXiv250818082T}, and population-informed samples, the majority of BBHs have effective spins predominantly occupying the positive region. Although the effective spins are slightly changed after reweighed by our population model, just a small fraction of events are shifted toward more positive values. We thus conclude that parameter correlations may modestly influence the quantitative outcomes of our analysis, but they do not alter the qualitative conclusions of this work.

We also note that there are several simulations about end-to-end injection and recovery of the mock populations of BBHs, which show that the effective-spin distribution can always be well recovered, see e.g., \citet{2021ApJ...922L...5C}, \citet{2024PhRvD.109j4036M} and also our previous work \citep{2025ApJ...987...65L}
In summary, the apparent alignment of spins we observe is physically real and not an artifact of systematic errors.
}

\section{Impact of likelihood uncertainty}\label{sec:var}
\begin{figure}
	\centering  
\includegraphics[width=0.9\linewidth]{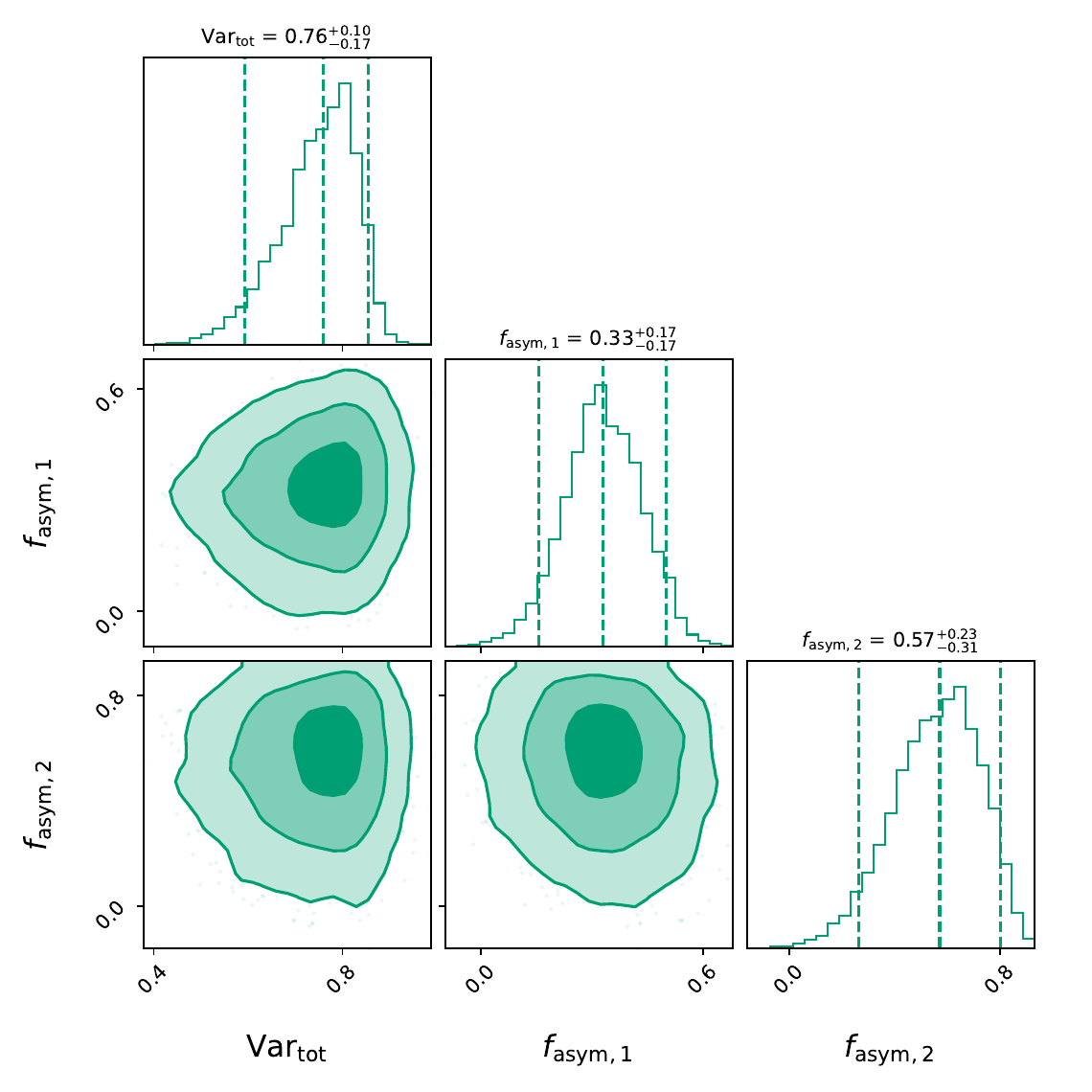}
\caption{Distribution of the likelihood uncertainty as well as the asymmetry fractions of the two subpopulations. The likelihood uncertainty is not correlated with the asymmetry fractions, suggesting that our main conclusion is solid.}
\label{app:var}
\end{figure}

\begin{figure*}
	\centering  
\includegraphics[width=0.98\linewidth]{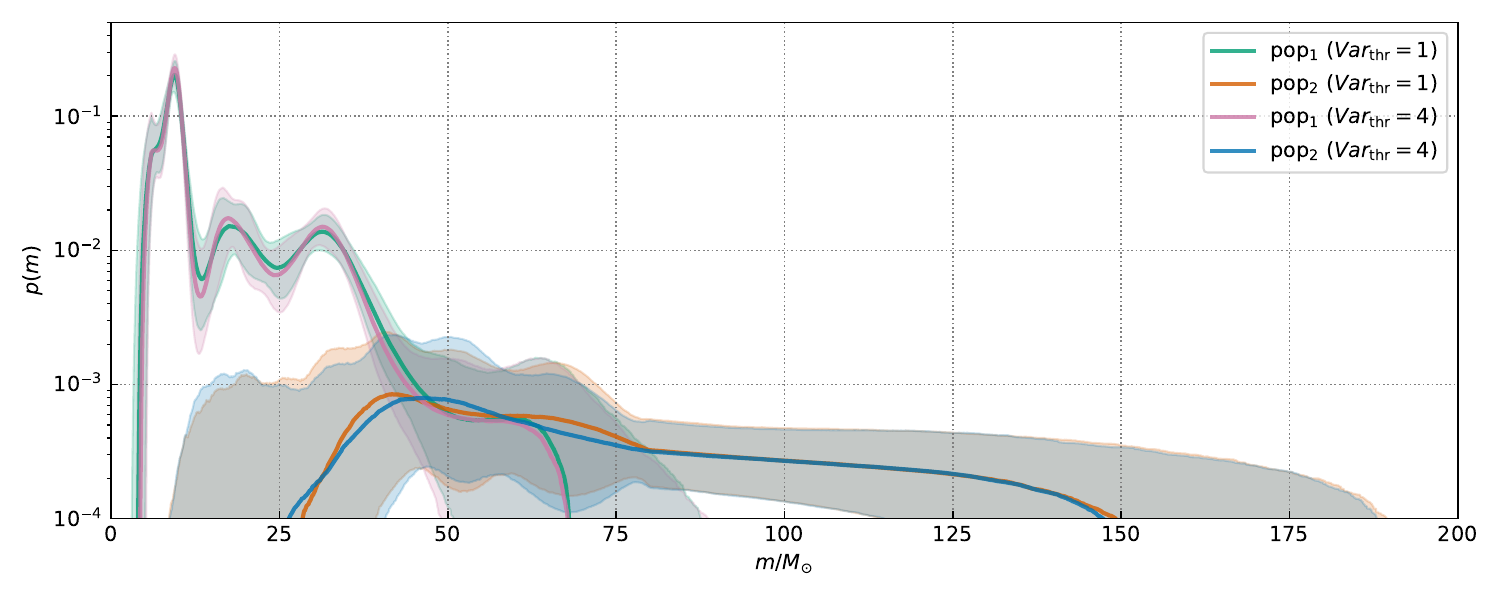}
\hspace*{.4cm}
\includegraphics[width=0.95\linewidth]{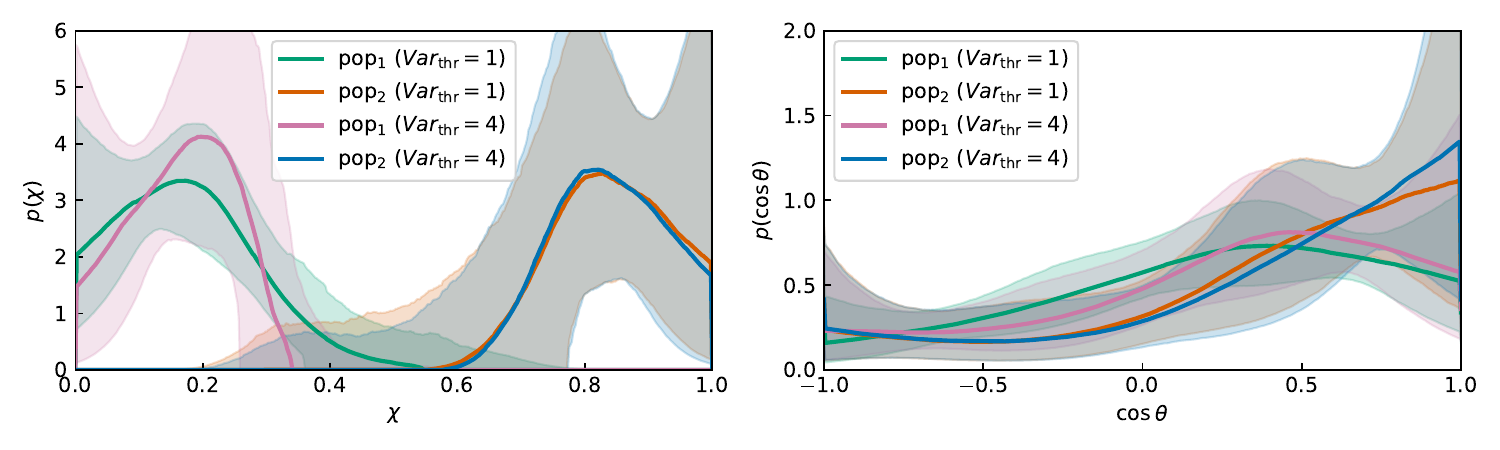}
\caption{Comparison of the results inferred with $Var_{\rm thr}=1$ and  $Var_{\rm thr}=4$. The mass and spin distributions are well consistent with each other in these two scenarios, demonstrating the robustness of our main conclusion. 
}
\label{app:compare_var}
\end{figure*}

Monte Carlo uncertainty in the likelihood evaluation may influence the hierarchical inference \citep{2023MNRAS.526.3495T}. To check whether such uncertainty has impact on our results, we plot the distribution of likelihood uncertainty as well as the asymmetry fractions of the two subpopulations in Figure~\ref{app:var}, and find out that the likelihood uncertainty has no tight correlation with the asymmetry fraction of the $\cos\theta$ distributions for the two subpopulations. This means that our main conclusion holds against the likelihood uncertainty.
We have also repeated our analysis with a more permissive cutoff $Var_{\rm thr}=4$, and compared to the results in the main text, i.e., inferred with  $Var_{\rm thr}=1$. It turns out that the mass and spin distributions are well consistent between the two results (see Figure~\ref{app:compare_var}), demonstrating that the main finding of this work, i.e., the aligned-spin fraction for the high-spin subpopulation is robust.

\end{document}